\documentclass[sigplan,screen]{acmart}

\copyrightyear{2025}
\acmYear{2025}
\setcopyright{cc}
\setcctype{by}
\acmConference[ASPLOS '25]{Proceedings of the 30th ACM International Conference on Architectural Support for Programming Languages and Operating Systems, Volume 2}{March 30-April 3, 2025}{Rotterdam, Netherlands}
\acmBooktitle{Proceedings of the 30th ACM International Conference on Architectural Support for Programming Languages and Operating Systems, Volume 2 (ASPLOS '25), March 30-April 3, 2025, Rotterdam, Netherlands}\acmDOI{10.1145/3676641.3716281}
\acmISBN{979-8-4007-1079-7/2025/03}

\usepackage[]{hyperref}

\usepackage{booktabs}
\usepackage[frozencache]{minted}
\usepackage{makecell}
\usepackage{framed}
\colorlet{shadecolor}{yellow!20}
\usepackage{soul}
\soulregister\cite7 
\soulregister\fixme1
\soulregister\ref1
\soulregister\texttt1
\soulregister\subsubsection1
\soulregister{\hl}{1}
\usepackage{xcolor}

\definecolor{codegreen}{rgb}{0,0.6,0}
\definecolor{codegray}{rgb}{0.5,0.5,0.5}
\definecolor{codepurple}{rgb}{0.58,0,0.82}
\definecolor{backcolour}{rgb}{0.95,0.95,0.92}
\usepackage{listings}
\lstdefinestyle{mystyle}{
    commentstyle=\color{codegreen},
    keywordstyle=\color{magenta},
    numberstyle=\tiny\color{codegray},
    stringstyle=\color{codepurple},
    basicstyle=\linespread{0.8}\ttfamily\footnotesize,
    breakatwhitespace=false,         
    breaklines=true,                 
    captionpos=b,                    
    keepspaces=true,                 
    showspaces=false,                
    showstringspaces=false,
    showtabs=false,                  
    tabsize=2
}
\newcommand{\fixme}[1]{{\color{red} FIXME: {#1}}}

\definecolor{skyblue}{RGB}{160, 250, 250}

\DeclareRobustCommand{\hll}[1]{{#1}}

\begin{document}

\title{Virgo: Cluster-level Matrix Unit Integration in GPUs for Scalability
and Energy Efficiency}

\author{Hansung Kim}
\authornote{Both authors contributed equally to this work.}
\orcid{0000-0002-0080-0548}
\affiliation{%
    \institution{University of California, Berkeley}
    \city{Berkeley}
    \state{California}
    \country{USA}
}
\email{hansung_kim@berkeley.edu}

\author{Ruohan Richard Yan}
\authornotemark[1]
\orcid{0009-0001-3695-6499}
\affiliation{%
    \institution{University of California, Berkeley}
    \city{Berkeley}
    \state{California}
    \country{USA}
}
\email{yrh@berkeley.edu}

\author{Joshua You}
\orcid{0000-0002-6317-6690}
\affiliation{%
    \institution{University of California, Berkeley}
    \city{Berkeley}
    \state{California}
    \country{USA}
}
\email{jyou12@berkeley.edu}

\author{Tieliang Vamber Yang}
\authornote{Work done while at UC Berkeley.}
\orcid{0009-0008-9316-689X}
\affiliation{%
    \institution{NVIDIA Corporation}
    \city{Santa Clara}
    \state{California}
    \country{USA}
}
\email{vambery@nvidia.com}

\author{Yakun Sophia Shao}
\orcid{0000-0003-1811-5407}
\affiliation{%
    \institution{University of California, Berkeley}
    \city{Berkeley}
    \state{California}
    \country{USA}
}
\email{ysshao@berkeley.edu}

\renewcommand{\shortauthors}{Hansung Kim, Ruohan Richard Yan, Joshua You, Tieliang Vamber Yang, Yakun Sophia Shao}



\begin{abstract}
Modern GPUs incorporate specialized matrix units such as Tensor Cores to
accelerate GEMM operations, which are central to deep learning workloads.
However, existing matrix unit designs are tightly coupled to the SIMT core,
restricting operation size due to register file capacity and bandwidth
constraints.
Such a limitation in scalability makes it difficult to simultaneously improve
compute throughput and energy efficiency in GPUs.


To address this challenge, we propose \emph{Virgo}, a GPU microarchitecture that
integrates dedicated matrix units at the \emph{SIMT core cluster} level.
By decoupling the matrix unit from the SIMT core, Virgo eliminates
scalability constraints imposed by the core microarchitecture.
Consequently, \hll{Virgo increases operation granularity at the hardware
level, reducing energy overhead from core instruction processing.
Physical disaggregation also enables a unified matrix unit design and
offloading both operand and accumulator accesses from the register file,
improving data reuse and energy efficiency.
}
Furthermore, this disaggregation supports efficient concurrent execution of the
SIMT core and matrix unit, optimizing mapping for fused DNN workloads.
%
%
Our evaluations using synthesizable RTL demonstrate that Virgo
achieves \hll{67.3\% and 24.2\% reduction in on-chip active power consumption,
  compared to the baseline Ampere-style and Hopper-style core-coupled designs.}

\end{abstract}

\maketitle 

\section{Introduction}

\begin{figure}[t]
  \centering
  \includegraphics[width=\linewidth]{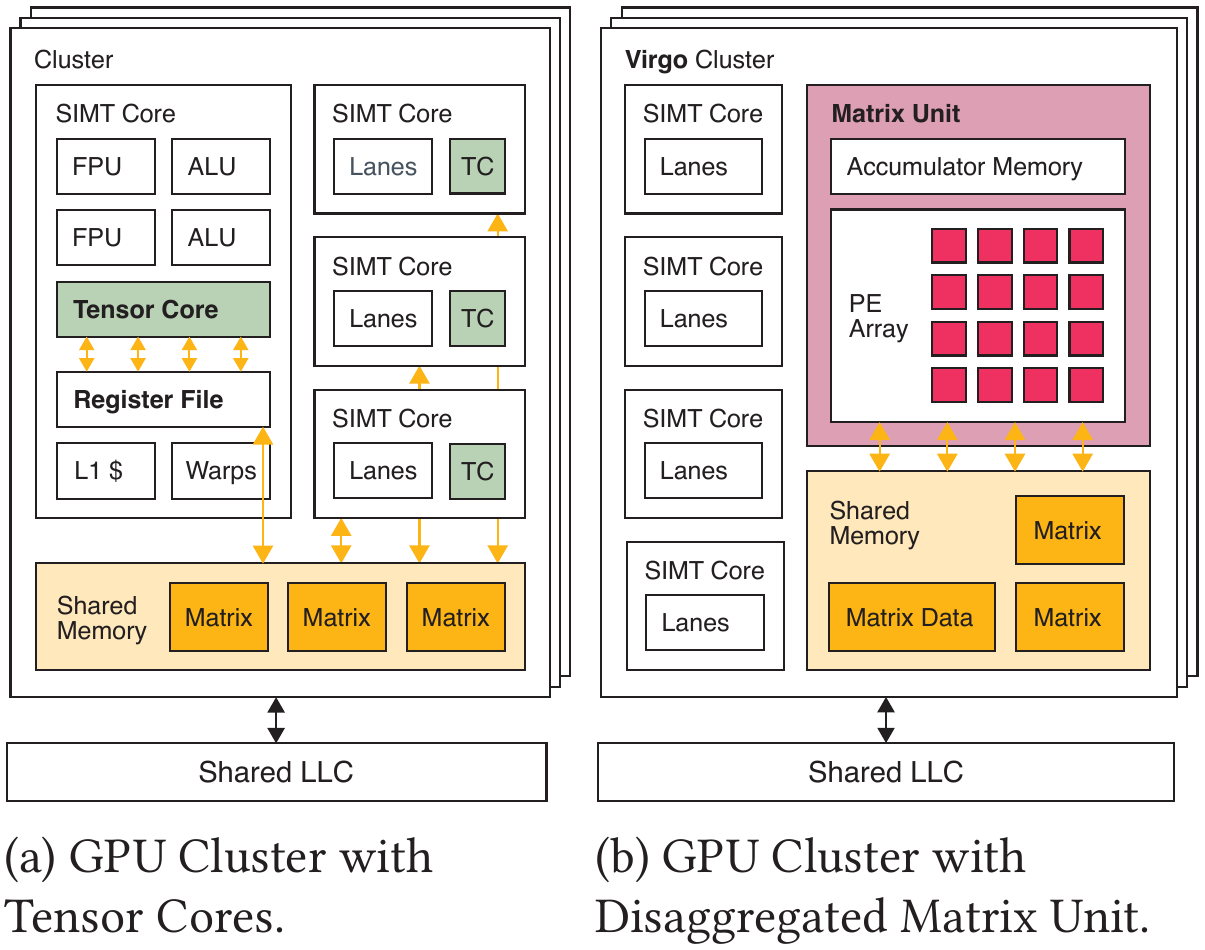}
  \caption{Overview of (a) today's GPU architecture with tightly coupled integration of Tensor Cores (TC), compared to (b) Virgo's cluster-level integration of
  matrix units.}
  \label{fig:overview}
\end{figure}


In recent years, the computational capability of GPUs has surged at an
unprecedented rate, driven by the demand of emerging large-scale deep learning
applications such as large language models~\cite{brown2020language,
sevilla2022compute, davies2024journey}.
To better meet these application demands, modern GPU architectures have increasingly incorporated specialized matrix
units, such as NVIDIA Tensor Cores~\cite{nvidia2017volta} and AMD Matrix
Cores~\cite{amd2021cdna2}.
These units accelerate GEMM operations in deep learning workloads with significantly higher compute efficiency than the traditional general-purpose SIMD units.

As the demand for higher compute capabilities grows, so too has the
scale of integration of dedicated matrix units: the number of FLOPS achieved by
Tensor Cores has increased eight-fold from Volta to Hopper, over the past five
years~\cite{nvidia2017volta,
nvidia2024hopper}.
Beyond FLOPS, power and energy have become increasingly important concerns for modern GPU workloads.
Namely, deep learning applications are known to be highly
energy-intensive workloads on GPUs~\cite{hodak2019towards}.
GPUs are also often power-limited; datacenters over-provision GPU resources, leading to frequent throttling to meet the power budget, which in turn compromises
performance~\cite{bridges2016understanding, li2022ai, patel2023towards,
zhao2023sustainable, luo2024benchmarking}.


However, it is increasingly challenging to simultaneously meet the demand for
higher FLOPS and better energy efficiency, due to the \emph{tight coupling}
between the matrix units and the GPU SIMT cores.
Matrix units are typically integrated into the SIMT core pipeline as specialized
functional units, receiving data through the
register file via the standard instruction datapath.
This tightly coupled integration faces significant limitations in terms of
operation size due to the capacity and bandwidth constraints of the register
file.
Consequently, core-coupled matrix units
support fine-grained operations, commonly with tile sizes such as 16 $\times$
8~\cite{sun2022dissecting, schieffer2024characterizing}. Such a small
granularity of operation not only limits data reuse, but also requires
processing a large number of instructions in the core pipeline, consuming
substantial energy and power in instruction scheduling and address generation,
rather than in actual computation.

To address these challenges, we propose Virgo, a novel GPU architecture that integrates dedicated matrix units at the \emph{cluster} level (Figure~\ref{fig:overview}).
The cornerstone of Virgo is the physical disaggregation of the matrix unit from
the core microarchitecture, eliminating scalability constraints 
and increasing matrix operation granularity.
\hll{Virgo improves system-level energy efficiency through reducing
instruction processing, eliminating redundant register file accesses, and
increasing data reuse.}
Key contributions of our work are: 

\begin{itemize}
\item{We propose a novel cluster-level matrix unit integration methodology for
GPUs that enhances scalability and efficiency by disaggregating the accelerator from the SIMT core.}
\item{We fully implement the GPU designs featuring not only the proposed
  cluster-level integration, but also \hll{baseline core-coupled integrations
    that model Volta, Ampere and Hopper Tensor Cores, in synthesizable RTL.}
    We develop the corresponding software programming interface as well. Virgo is fully open sourced.
    \footnote{RTL: \url{https://github.com/ucb-bar/virgo},
    kernels: \url{https://github.com/ucb-bar/virgo-kernels}.}}



  \item{\hll{We demonstrate that Virgo, when synthesized using a commercial 16nm
    process, significantly reduces active power consumption by 67.3\% and 24.2\% compared to the
    Ampere and Hopper-style core-coupled baselines, and energy consumption
    by 80.3\% and 32.5\%.}}

\end{itemize}

\section{Background and Motivation}

This section provides an overview of the GPU and Tensor Core microarchitectures, highlighting the scalability and efficiency limitations caused by the tight coupling of the matrix unit to the SIMT core.
This discussion motivates Virgo's
key design decision to integrate matrix units at the cluster level, addressing architectural bottlenecks and aligning with the increasing demands of modern GPU applications.


\subsection{GPU Cluster Microarchitecture}

Figure~\ref{fig:overview}(a) shows the modern GPU microarchitecture, featuring multiple
clusters of SIMT cores connected to a shared last-level cache.
Each SIMT core consists of a warp scheduler, a register file, and tightly-coupled execution
units like Tensor Cores. 
Within each cluster, SIMT cores are
interconnected via a cluster-level network-on-chip to the \emph{shared memory},
a software-managed scratchpad.


The clustered organization offers several benefits.
First, it increases hardware parallelism by allowing multiple cores in a cluster
to execute independent warp instructions simultaneously.
Second, it enables increased data sharing within the kernel, as a greater number
of threads across the SIMT cores can share data through the shared memory.
This cluster-based architecture is widely adopted in the industry, known as \emph{Streaming Multiprocessors} in NVIDIA
GPUs~\cite{nvidia2024hopper}, \emph{Compute Units} in AMD CDNA
architecture~\cite{amd2021cdna2} and \emph{$X^e$-cores} in Intel $X^e$-HPG
architecture~\cite{intel2022xe}.


Importantly, the cluster serves as the hardware unit to which a \emph{thread block} or a
\emph{workgroup} is assigned,
which is a group of SIMT threads that provides software primitives for shared
memory access and barrier synchronization~\cite{gilman2021demystifying}. With
modern workloads like large language models (LLMs) requiring increasingly large
GEMM operations, there is a growing need for improved data reuse at the shared
memory level. This necessity motivates the rationale for integrating matrix
units at the cluster level, a key principle that guides the development of the
Virgo system.

\subsection{State-of-the-art Tensor Core Integration}

To meet the rapidly increasing compute demand of 
deep learning applications, NVIDIA introduced dedicated matrix units known as
\emph{Tensor Cores} in the Volta architecture~\cite{nvidia2017volta}.
Tensor Core consists of multiple SIMD-parallel dot product units designed for high-throughput 
multiply-add operations in matrix multiplication~\cite{raihan2019modeling}. 
As shown in Figure~\ref{fig:overview}(a),
Tensor Cores receive matrix operands directly through the register file of
the SIMT core, similar to other execution units such as the floating-point unit
and the load/store unit. Essentially, today's Tensor Core is a specialized
execution unit for matrix multiplication that is tightly coupled into the
execution pipeline of the SIMT core. 
Prior efforts have focused on improving the Tensor Core
design in sparsity support~\cite{wang2021dual, huang2023rm, fan2024dtc} and
generalized operations~\cite{zhang2022simd2, sung2023mad}, while maintaining the core-coupled integration.


\begin{table}[t]
\centering
\begin{tabular}{c|cccc}
\toprule
GPU & V100 & A100 & H100 \\
\midrule
Architecture & Volta & Ampere & Hopper \\
\makecell{Tensor FP16 TFLOPS} & 1x & 2.5x & 7.9x \\
\makecell{CUDA FP32 TFLOPS} & 1x & 1.2x & 4.3x \\
\# of Tensor Cores (TC) & 1x & 0.7x & 0.8x \\
MACs per TC & \textbf{64} & \textbf{256} & \textbf{512} \\

\midrule

\makecell{\hll{Register usage (max. 255)}} & \hll{224} & \hll{221} & \hll{168} \\
\makecell{\hll{Warp occupancy}} & \hll{12.5\%} & \hll{10.0\%} & \hll{14.1\%} \\

\bottomrule

\end{tabular}

\caption{Scaling trends of the compute capabilities of NVIDIA datacenter GPUs
  across generations, and \hll{occupancy statistics of CUTLASS GEMM kernels on
  them.} The multiply-accumulate (MAC) units per Tensor Core is analytically
  estimated from FLOPS using the clock frequency.
  For the kernel statistics, we profile \texttt{s884gemm\_f16*},
  \texttt{s16816gemm\_f16*} and \texttt{sm90*\_s64x128x16gemm\_f16*} for
  V100/A100/H100, and compute the average across five kernel parameter
  combinations with the highest FLOPS for each architecture.}
\label{table:tensor-core-scaling}
\end{table}


Notably, Tensor Core has been rapidly increasing in its compute capabilities.
Table~\ref{table:tensor-core-scaling} outlines the evolution of NVIDIA datacenter GPUs across generations~\cite{nvidia2017volta,
nvidia2022amperedata, nvidia2024hopper}.
The FP16 throughput of Tensor Cores has improved at a rate surpassing that of FP32 throughput in CUDA cores.
This increase in throughput is not due to an increase in the number of Tensor Cores, but rather to each Tensor Core
instance growing larger in size.
This trend underscores the need to scale up the individual capabilities of the
matrix units for future generations of GPUs. 



\subsection{Limitations of the Core-Coupled Approach} \label{sec:core-coupled-limitations}
However, the tightly core-coupled nature of today's Tensor Core design
poses significant challenges to further scaling of the matrix unit.
First, modern GEMM kernels 
generate high
\emph{register pressure} as they require extensive use of register file space to
store multiple input and accumulator data~\cite{kerr2017cutlass}.
This issue is compounded in kernels that utilize dedicated matrix units, where the
entire input and accumulator matrix tiles must be stored within the register
file, significantly increasing the register usage.
As a result, this leads to decreased kernel occupancy or frequent register spills
to stack memory in Tensor Core-accelerated GEMM kernels, as reported in previous studies~\cite{tan2011fast, wang2016optimizing,
zhang2017understanding, yan2020demystifying, bikshandi2023case}.
\hll{
Our characterization of CUTLASS GEMM kernels~\cite{kerr2017cutlass}
 confirms this as well, where high register usage significantly limits warp occupancy of
the kernel (Table~{\ref{table:tensor-core-scaling}}).
}
Efforts like INTERPRET~\cite{kwak2023interpret} and Duplo~\cite{kim2020duplo} are developed to reduce duplicated data in the register file, alleviating capacity constraints.


Furthermore, the tight coupling to the SIMT core's register file imposes not only
\emph{capacity} constraints, but also \emph{bandwidth} constraints.
According to \cite{raihan2019modeling}, the register read bandwidth available in
the Volta architecture is maximally utilized during Tensor Core operations.
Scaling up the operation size requires increasing register file bandwidth to
deliver larger operand data to the Tensor Core, introducing significant design
challenges.

Due to register file constraints, core-coupled matrix unit designs typically support \emph{fine-grained} operations. 
For example, the NVIDIA Tensor Cores handle tile sizes of 16 $\times$ 8 $\times$ 16
for FP16 input operands in the MMA instruction~\cite{nvidia2024ptx};
similarly, AMD CDNA2 Matrix Cores supports sizes of 16 $\times$ 16 $\times$ 16 and
32 $\times$ 32 $\times$ 8 in the MFMA
instruction~\cite{schieffer2024characterizing}.  
These small tile sizes incur additional data transfers from the backing memory
due to limited reuse.  Additionally, they require processing a large number of
instructions and computing extra tile memory addresses to complete the GEMM
operation, leading to significant energy and power consumptions in hardware
components beyond the matrix unit, including the instruction scheduling logic,
ALU, and the register file.






\subsection{Recent Advances and Remaining Challenges} \label{sec:background-recent}

\hll{
Recent industry GPU designs have made advancements in addressing the efficiency
and scalability challenges of the core-coupled approach described above.
Namely, NVIDIA Ampere Tensor Core introduces Asynchronous Data
Copy~\cite{nvidia2024ampere}, which adds hardware acceleration for transferring
data data directly from the global memory to the shared memory, bypassing the
core's register file.
By offloading data copy operations from the core, resource usage within the core
pipeline---such as register files and instruction slots---is reduced, enabling more
efficient allocation of such resources to matrix operations.
Additionally, this offloading
enables asynchronous overlapping of data movement and compute, improving
efficiency compared relying solely on warp concurrency.  In our profiling of
CUTLASS, we observe 36\% less warp stalls in the async-copy-enabled GEMM
kernels, which led to 22\% increase in issue slot utilization.}

Building upon Ampere, NVIDIA's Hopper Tensor Core, introduced in 2022, further
alleviates register pressure with the introduction of the \texttt{wgmma}
operation.  This new mode of operation enables the Tensor Cores to read matrix
operands directly from shared memory, reducing dependency on the register
file~\cite{luo2024benchmarking, nvidia2024hopper}. This new operand delivery
method mitigates the need to store the matrix data in the register file,
alleviating the scalability limitation.  This is evidenced by the significant
decrease in register usage that enabled further scale-up of MACs in Hopper, shown
in Table~\ref{table:tensor-core-scaling}.

However, the Hopper Tensor Core does not fully solve the register pressure
issue, and energy efficiency remains a concern. The \verb|wgmma| instruction,
while reducing some register load, still accumulates the partial sum matrices
back to the register file~\cite{nvidia2024ptx}, generating significant register
pressure (Table~\ref{table:tensor-core-scaling}).
Furthermore, the Tensor Core still remains physically coupled to the SIMT
core~\cite{luo2024benchmarking, nvidia2024hopper}, preventing data sharing
across matrix units that could otherwise be achieved in a large, unified matrix unit design.

\hll{Virgo fundamentally address the scalability and energy efficiency
challenges by completely disaggregating the matrix unit from the SIMT cores,
establishing it as a separate instance at the cluster level.  To the best of our
knowledge, Virgo is the first effort to integrate a matrix unit at the core cluster.
Through disaggregation, Virgo completely bypasses the core's register file for
operand and accumulator storage, eliminating its scalability constraints.  It
also reduces energy consumption in the register file, by offloading data access
directly to the shared memory or a separate, energy-efficient accumulator
memory.
Finally, the disaggregation also enables a unified matrix unit
design, which maximizes data reuse across processing elements compared to
separate, per-core configuration.
}
%

\subsection{Terminology and Modeling} \label{sec:terminology}

\hll{
To facilitate further discussion, we define a terminology that corresponds to the
integration methods described above.
This classification allows us to separate different design concerns and
facilitate a fair evaluation of our design.
Section~\ref{sec:methodology} details how we model each of these design points.
}

\begin{itemize}
  \item{\hll{\textbf{Tightly-Coupled Matrix Unit} refers to the traditional
    design where the matrix unit is integrated within the SIMT core,
    heavily relying on the register file for both operand and result data
    delivery.  The NVIDIA Volta Tensor Core is an example of this design.}}
  \item{\hll{The Tightly-Coupled Matrix Unit can be extended with a
    \textbf{dedicated DMA}, enabling more efficient data delivery from off-chip
    to on-chip memory.  This configuration corresponds to Ampere,
    where we assume that Ampere features a separate DMA engine similar to Hopper
    in order to support Asynchronous Data Copy instructions.}}
  \item{\hll{\textbf{Operand-Decoupled Matrix Unit} allows the matrix unit
    to directly source operand data from the shared memory, but the unit itself
    remains to be part of the SIMT core.  This configuration resembles Hopper.}}
  \item{\hll{\textbf{Physically Disaggregated Matrix Unit} refers to Virgo's
    design, where the matrix unit is completely separated from the SIMT core
    and externally integrated at the cluster, enhancing data reuse and
    scalability.}}
\end{itemize}

\hll{
For brevity, we also refer to the above terms as \emph{Volta-style},
\emph{Ampere-style}, \emph{Hopper-style matrix units} and \emph{Virgo}.
}

\newpage

\section{Virgo Microarchitecture} \label{sec:microarchitecture}

\begin{figure}[t]
  \centering
  \includegraphics[width=\linewidth]{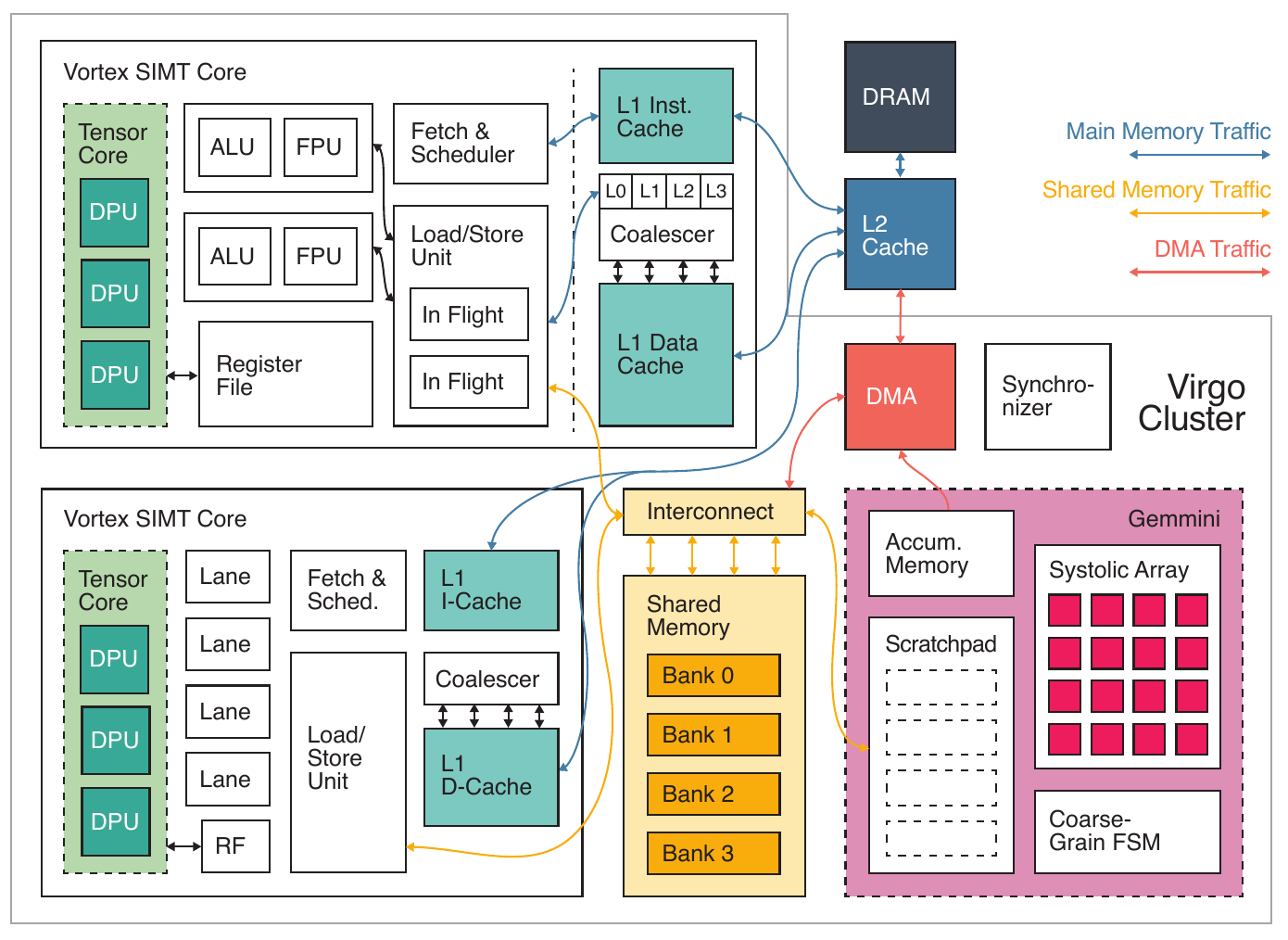}
  \caption{A high level overview of the Virgo microarchitecture.  The
  Gemmini-based matrix unit is disaggregated out from the SIMT cores into a
  separate unit in the cluster.  Dashed lines indicate optionally instantiated
  modules for evaluation, such as the core-coupled Tensor Core.}
  \label{fig:cluster-uarch}
\end{figure}

Virgo aims to improve scalability and energy efficiency by disaggregating the
dedicated matrix unit from the core as separate hardware at the cluster level,
as depicted in Figure~\ref{fig:cluster-uarch}.

To realize this design, we need to solve several microarchitectural challenges:
(1) establishing a command interface through which the core and the unit can
communicate, (2) designing a cluster-local shared memory interconnect that
efficiently handles concurrent accesses from the core and the matrix unit, (3)
facilitating a more efficient matrix data movement via a dedicated DMA engine.
and (4) implementing a synchronization mechanism across the cores and the matrix
unit.
We discuss these components in more detail in the following subsections.



Before we begin, we mention key open-source hardware infrastructures that we
leverage to realize our design at the RTL.  First, we use
Vortex~\cite{tine2021vortex}, an open-source GPGPU implementation in
SystemVerilog that enables the full stack of both hardware and kernel
development by extending the RISC-V ISA. Second, we use
Gemmini~\cite{gemmini-dac}, a systolic array generator, to
generate the cluster-based matrix unit IP for Virgo\footnote{The name
\emph{Virgo} is inspired by the use of \emph{Vortex} and \emph{Gemmini} to achieve
\emph{cluster}-level integration.}. We describe in further detail how we modify these IPs
to establish a full GPU system in Section~\ref{sec:methodology}.



\subsection{Command Interface to the Matrix Unit} \label{sec:gemmini-uarch}


We facilitate the cluster-local interconnect to establish an efficient
memory-mapped IO-based command interface between the core and the matrix unit.
The MMIO interface is advantageous in our design because it does not require
modification of the ISA and the core microarchitecture: At the kernel,
the core controls the matrix unit simply by issuing regular loads and stores to
a specific memory address region.  At the same time, the core still benefits
from low-latency access to the memory-mapped registers that are interconnected
to the lightweight, cluster-local NoC.

In order to establish the MMIO interface, we replace Gemmini's original RoCC
interface~\cite{asanovic2016rocket} with memory-mapped control registers that
the core can access via the shared memory address space.
To synchronize the core threads in relation to Gemmini operations, we use a
software routine that polls a memory-mapped register that indicates the
accelerator's busy state.
When performing GEMM, a hardware FSM exists in Gemmini to automatically iterate
through the $i$, $j$ and $k$ dimensions, allowing a single invocation to fully
compute on matrix sizes larger than the systolic array dimensions. The FSM
retrieves operands from the shared memory, controls systolic array operation,
and writes the result matrix to the accumulator memory.



\subsection{Memory System}

The main challenges of Virgo's memory system involves handling concurrent,
heterogeneous accesses to the shared memory from both the core and the matrix
unit, and achieving high effective bandwidth in fetching matrix data from the
global memory.  The following subsections discuss each of these challenges.


\subsubsection{Shared memory} \label{sec:uarch-shared-memory}

The shared memory must efficiently support specific access patterns from both
the SIMT cores and the matrix units.  Namely:

\begin{itemize}
\item{The Gemmini-based matrix unit generates wide accesses of size \texttt{4n} bytes,
where \texttt{n} is the systolic array dimension.  On the other hand,
the individual lanes of the SIMT core make narrower 4-byte requests.}
\item{The SIMT cores and the matrix unit should be able to concurrently access two different
matrices without serialization.  This is necessary when they operate in a
producer-consumer relationship through double buffering in the
memory, as described in Section~\ref{sec:software-pipeling}.}
\end{itemize}



\begin{figure}[t]
  \centering
  \includegraphics[width=\linewidth]{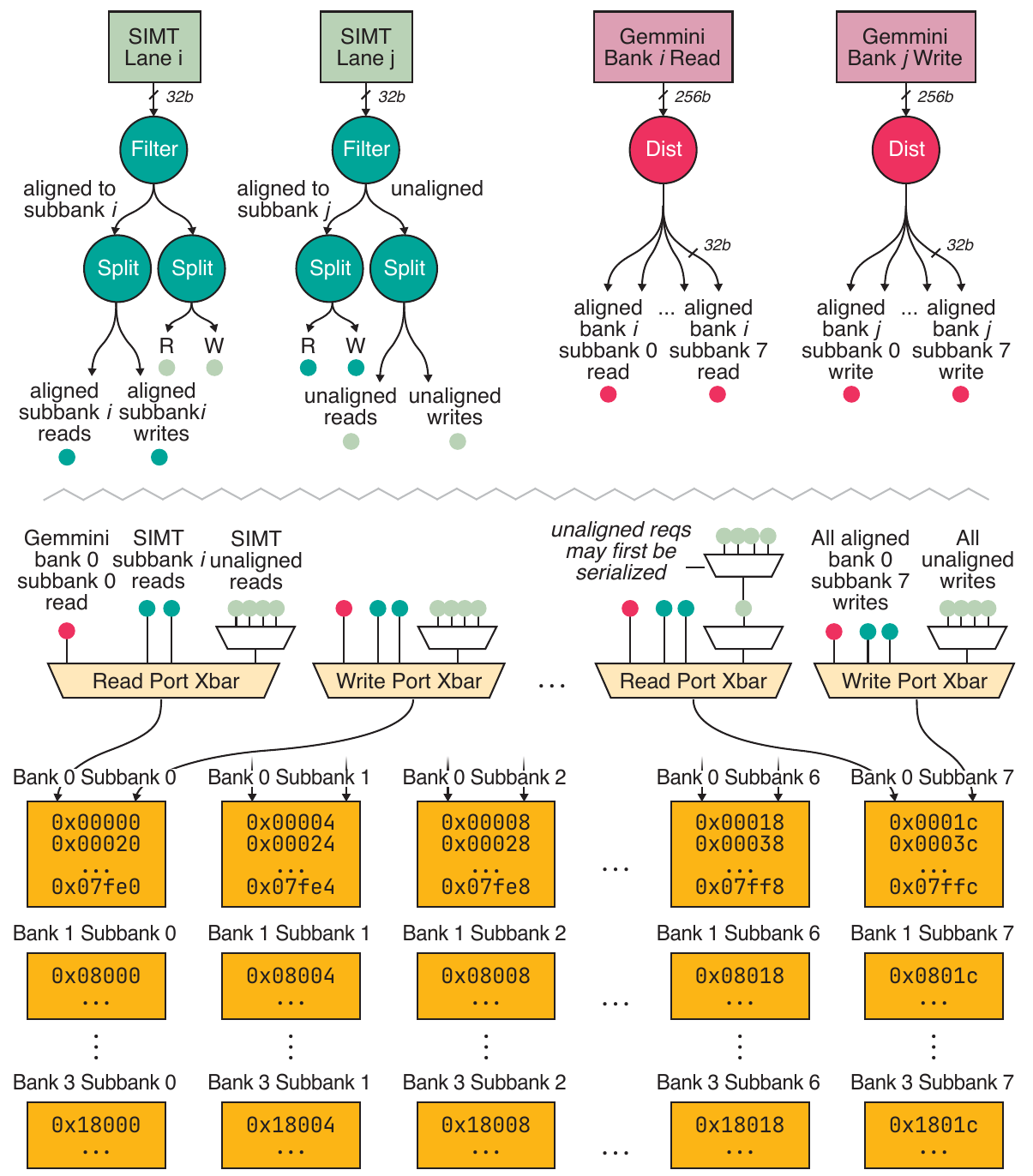}
  \caption{The Virgo shared memory system.
  }
  \label{fig:smem-system}
\end{figure}

It is also important to keep the hardware complexity of the interconnect
manageable while supporting the above access patterns, given that there are
many SIMT lanes in the cluster.
To tackle these challenges, we make the following key design choices, depicted
in Figure~\ref{fig:smem-system}:

\textbf{Two-dimensional banking.}  To accommodate the aforementioned
different-sized accesses from the core and matrix units while incurring minimal
bank conflicts, we partition shared memory address space in two dimensions:
banks and subbanks.
This enables parallelism of memory accesses across subbanks, i.e. from SIMT
lanes within a core, as well as banks, i.e. from both matrix unit and cores.

\textbf{Unified request sizes.}  We divide a wide matrix unit request into
multiple word-sized sub-requests, and distribute them across the subbanks within
a single bank.
If requests arrive at the same bank from both cores and the matrix unit, the
interconnect prioritizes the wider matrix requests, serving all sub-requests in
the same cycle, which allows the matrix unit to run at full throughput.

\textbf{Separate read and write paths.} The requests originated from the core
uses the same channel for reads and write; the shared memory interconnect routes
them to separate channels to avoid conflicts and merges the responses on the
return direction.
This better supports the aforementioned producer-consumer relationship
scenarios.

\textbf{Area optimization for unaligned SIMT accesses.}
Interconnecting every core's SIMT lanes into every subbank via a crossbar
results in a high area cost. Instead, we selectively filter out SIMT memory
requests whose addresses are not aligned to a word boundary, and serialize them
into one lane before entering the crossbar. This optimization makes minimal
impact to the evaluated kernels, as they frequently make word-aligned accesses
for moving matrix data.


The interconnect is implemented using the TileLink protocol~\cite{tilelink} and
Diplomacy~\cite{diplomacy} parameter negotiation framework, which enables
a flexible, parameterized design.
\hll{
We find that the interconnect design incurs minimal overhead.
Implemented and synthesized in RTL, the shared
memory accounts for 5.5\% of the area and 6.8\% of the power consumption in the
SoC.  Also, compared to a baseline design that does not support Gemmini
connection, the shared memory area only increases by 9.6\%.}

\subsubsection{Accumulator Memory} \label{sec:accum-mem}
Virgo's disaggregated matrix unit features a dedicated SRAM memory for storing
accumulator matrix data, also shown in Figure~\ref{fig:cluster-uarch}. Equipping
the matrix unit with a dedicated accumulator memory is an important design
choice we make for two reasons.

First, this design enables providing the matrix units with a large accumulator
data storage without impacting parallelism of the kernel. GPU's register file
space is privatized across different warps in the core, establishing the SIMT
abstraction.  Because core-coupled matrix units such as Tensor Cores accumulate
data and return the result also through the SIMT registers, scaling up the
matrix units increases per-warp register usage, and results in higher register
pressure and decreased occupancy of the kernel (see also
Table~\ref{table:tensor-core-scaling}). While lower occupancy does not always lead to lower
matrix FLOPS, it may affect the performance of fused DNN kernels where the
activation compute has to run with reduced thread-level parallelism. In
contrast, providing matrix units with access to a separate accumulator memory
without SIMT privatization decouples the issue of register pressure from the
size of matrix compute, and allows the hardware designer to scale up matrix
units without worrying its impact to the overall kernel performance.

%

Second, regular access patterns to the accumulator data enables a simpler and
more energy-efficient memory design.  Unlike the register file, which must
support divergent scatter-gather SIMT accesses, matrix units perform wide and
contiguous accesses to the accumulator data.  This enables a straightforward SRAM
implementation---specifically, the Virgo unit employs a single-banked SRAM,
reducing hardware complexity and lowering energy consumption per access. We
present quantitative results in Section~\ref{sec:gemm-power-energy}.


\subsubsection{Memory coalescer}
\emph{Memory coalescing}~\cite{davidson1994memory} plays a crucial role in GPU
memory systems by converting scalar SIMT memory accesses into wide, vectorized
accesses, and thereby achieving high \emph{effective} memory throughput. At the
onset of the project, Vortex had lacked hardware support for memory coalescing,
significantly limiting the memory bandwidth of SIMT load and store instructions.
To this end, we implemented a custom memory coalescing unit as shown in
Figure~\ref{fig:cluster-uarch}.  The coalescer monitors the core's SIMT memory
accesses at the core and L1 cache interface, and opportunistically merges them into
wider memory requests of the same size as the L1 cache line. The coalescer
significantly increased data delivery rate from the global memory (\emph{i.e.}
DRAM, L2 and L1) to the cores, and improved utilization of
the matrix units.  Note that the coalescer is used for the Tightly-Coupled,
Volta-style configuration (Section \ref{sec:terminology}), which utilizes SIMT
instructions for data delivery instead of a DMA.

\subsubsection{Efficient Data Movement with DMA}  \label{sec:dma}
We further facilitate efficient data delivery to the matrix units beyond what memory
coalescing provides by incorporating a dedicated DMA engine.
Hardware-accelerated data movement is becoming increasingly common in modern GPU designs.
The NVIDIA Ampere architecture introduces Asynchronous Data Copy to accelerate
direct memory transfers between global and shared
memory~\cite{nvidia2024ampere}, while the Hopper architecture features a
dedicated TMA engine that supports flexible address generation
modes~\cite{nvidia2024hopper}.
In our models of Ampere-style and Hopper-style Tensor Core systems, as well as
in Virgo, we include a MMIO-programmable DMA engine in the cluster dedicated to
issuing direct memory accesses between global memory and shared memory.
Additionally in Virgo, the same DMA is capable of drawing from the matrix unit's
accumulator memory as well, depicted in Figure~\ref{fig:cluster-uarch}.



\subsection{Cluster-wide Synchronization} \label{sec:cluster-wide-sync}

As will be discussed in Section~\ref{sec:programming-model}, the SIMT cores in a
cluster work together collaboratively to either move matrix data to the matrix
unit, or do post-processing computation on the resulting matrix from the unit.
This requires an efficient synchronization mechanism across the cores to be
implemented at the entire cluster level.


To this end, we design a lightweight synchronizer module in the
cluster that interfaces with the warp scheduler of each SIMT core.  When a
designated set of warps reach a barrier instruction in the kernel, the warp
scheduler issues a barrier release request to the synchronizer.  The
synchronizer collects requests from other cores in the cluster, and replies once
receiving requests from all cores, ensuring every core participates in the
barrier.
We reuse Vortex's \verb|vx_bar| instruction to allow the programmer to specify
which warps participate in a particular barrier, and be able to use multiple
barriers in the kernel.

\section{Virgo Programming Model} \label{sec:programming-model}


As a result of architectural disaggregation from the core, the matrix unit in
Virgo operates as a separate thread of execution from the SIMT cores.  In this
section, we establish a programming model that enables the programmer to
efficiently coordinate the matrix unit and the SIMT cores, based on two
principles: (1) an asynchronous programming interface and (2) collaborative
execution of warps.


\subsection{Asynchronous Matrix Unit Interface} \label{sec:programming-async}

Virgo exposes an asynchronous programming interface to allow the programmer to
flexibly overlap SIMT core compute concurrently to the matrix operations. As
discussed in Section~\ref{sec:gemmini-uarch}, the Gemmini-based matrix unit in
Virgo exposes a memory-mapped IO interface to the core.  The accesses to the
MMIO interface are non-blocking, such that a single SIMT warp can kick off a
matrix unit operation and then continue to execute the next instructions in the
kernel without stalls. This enables a software-pipelining scheme where the warp
can overlap independent operations such as data movement for the next tile or
post-processed activation while the long-latency matrix operations occur.
We will detail how the programmer can exploit the instruction-level parallelism
through this interface in Sections \ref{sec:gemm-mapping} and
\ref{sec:flash-mapping}.


\hll{
An asynchronous matrix unit design is common in modern GPU architectures.
Notably, the NVIDIA Hopper Tensor Core~\cite{nvidia2024hopper} and
the AMD CDNA2 Matrix Core~\cite{amd2024matrix} support same-warp
co-execution of matrix and SIMT vector computations.
Without such a mechanism, the SIMT core relies solely on multithreading across
multiple warps to hide the latency of matrix operations, which requires high
warp occupancy to ensure the scheduler always has unblocked warps for issue.
In contrast, the asynchronous interface enables programmers to exploit
instruction-level parallelism within a single warp, providing additional
opportunities for latency hiding.  Virgo particularly benefits from this
interface, as its enhanced scalability supports larger tile sizes, which
introduce longer compute latencies to be overlapped.
}




\begin{figure}[t]
  \centering
  \includegraphics[width=\linewidth]{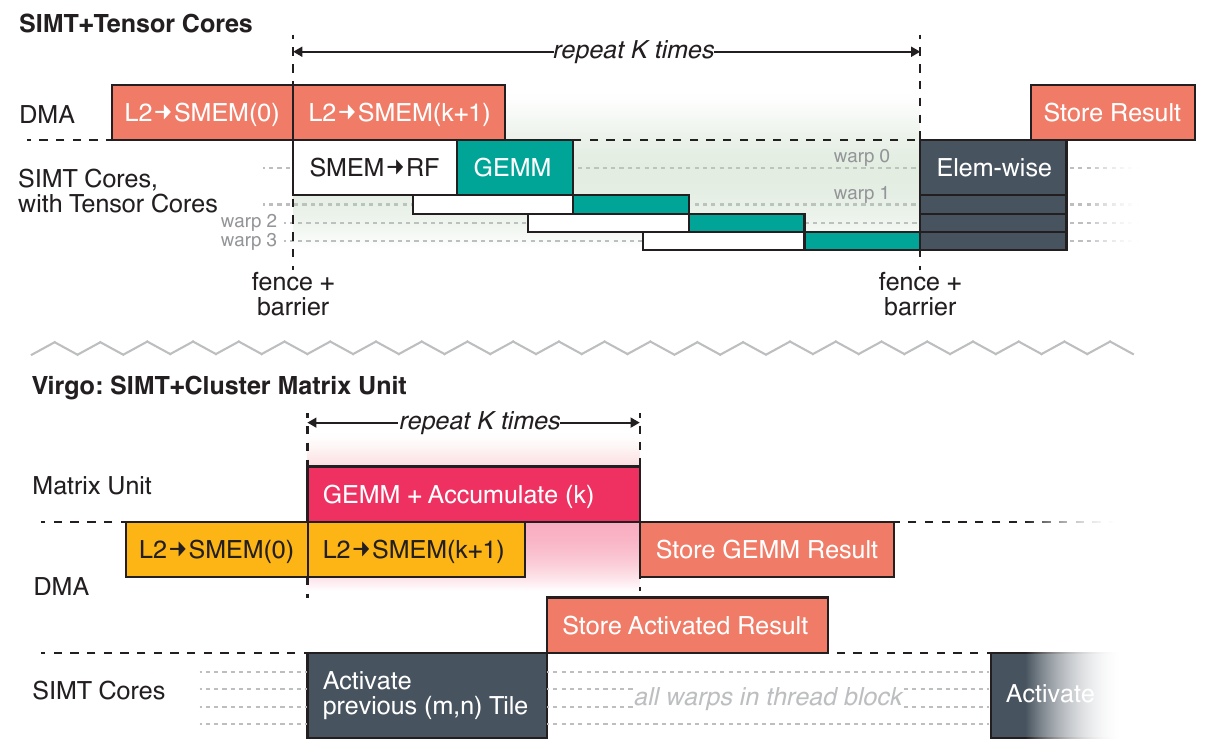}
  \caption{Execution in the SIMT core, matrix unit, and DMA during a GEMM
  operation on Virgo and Tensor Core-based designs.}
  \label{fig:pipeline}
\end{figure}

\subsection{Collaborative Execution of Warps}

GEMM operation on GPUs require close coordination between the core operation and
the dedicated matrix unit operation, where the core handles data movement
operations between the off-chip and on-chip memory, or post-processing compute
on the result matrix such as non-linear activations in a deep learning workload.
Therefore, the SIMT cores need to achieve adequate throughput that matches
compute throughput of the matrix unit. To this end, in Virgo, \emph{multiple}
warps collaborate to participate in a \emph{single} operation of the matrix
unit. This is depicted in Figure~\ref{fig:pipeline}, where all warps in the
thread block works on activating the previous (M,N) output tile computed by the
matrix unit operation. The collaborative execution of warps requires the
cluster-wide synchronization mechanism described in
Section~\ref{sec:cluster-wide-sync}, because the warps in the thread block may
be running in different cores in the cluster.

\subsection{API Design} \label{sec:programming-api}


\hll{
Virgo's application programming interface consists of two levels: A low-level
interface that directly exposes the initiation and synchronization mechanism
of the hardware, and a higher-level library that helps the programmer to
express GEMM computations.
The low-level API is composed of the following operations:}

\begin{itemize}
  \item{\hll{\texttt{virgo\_dma\_\{load,store\}} initiates an asynchronous
    DMA operation to copy the tile across the global memory, the shared memory
    and the accumulator memory inside the matrix unit.  It takes as argument
    the base addresses of the tiles, matrix dimensions, and their memory layout,
    which configures the DMA for correct addresses generation.}}
  \item{\hll{\texttt{virgo\_compute} kicks off the asynchronous matrix
    multiply-and-accumulate operation in the matrix unit, accessing the tile
    data from the shared memory.}}
  \item{\hll{\texttt{virgo\_fence} blocks the warp until all
    preceding asynchronous operations have completed and their results are
    visible to the programmer.}}
\end{itemize}

\hll{
As described in the previous section, these API calls are used in conjunction
with the cluster-wide synchronization barrier to ensure all participating cores
remain synchronized in relation to the matrix operation.

On top of the low-level API, we provide a library that constructs a full GEMM
kernel using tiled matrix multiplication.  The library consists of C++
templates that, given GEMM dimension constants, instantiates to a concrete
kernel specialized to each GEMM size by determining (M,N,K) loop iterations at
compile time. Using the library, the programmer can easily invoke a
device-level GEMM kernel without worrying about the details of mapping to the
matrix unit, or write a fused kernel that includes a GEMM component.

This API design is similar to existing GPU GEMM libraries such as Composable
Kernel~\cite{amd2024composable} or CUTLASS~\cite{kerr2017cutlass}, which
similarly construct a hierarchy of API layers on top of the
hardware-dependent, tile-level operators. 
One difference is that, in Virgo, the indivisible \emph{atom} of computation is
a threadblock- or workgroup-level matrix multiplication as a result of the
enlarged tile size, while the other libraries map to finer-grained operations at
the warp or warpgroup-level atoms.
Virgo's larger compute unit enables more optimal hardware scheduling decisions
while reducing instruction processing overhead within the core, as discussed in
as discussed in Section~{\ref{sec:gemm-power-energy}}. However, this approach
sacrifices flexibility for workloads that use smaller matrix dimensions.
}

\subsection{Mapping to the GEMM Kernel} \label{sec:gemm-mapping}

We now describe how the aforementioned programming model can be used to write an
optimized GEMM kernel.

\subsubsection{Thread block tiling}

We extend the well-established work partitioning scheme of GEMM kernels
on GPUs,
which employs tiling of the input matrices at \emph{multiple} levels to maximize
data reuse at the different memory hierarchy~\cite{kerr2017cutlass}.  Virgo's
matrix unit accelerates the first level of \emph{thread block} tiling, which
leverages parallelism across clusters and caches tiles at the shared memory. In
comparison, Tensor Cores accelerate the second level of \emph{warp} tiling.

In our configuration, the matrix unit exposes 128\texttimes64\texttimes128 as
the tile size of a single operation, which determines the thread block size.
Each thread block, spatially partitioned across the (M,N) output space,
completes the full GEMM by accumulating across the K dimension temporally in a
loop, shown in Figure~\ref{fig:pipeline}.  As the loop iterates, the Gemmini
matrix unit accumulates partial sum data onto its private accumulator memory,
which gets moved out and stored to the global memory at the end of the loop.
Then, the kernel moves on to the next (M,N) output tile that is allocated to
that thread block.
The kernel can launch multiple thread blocks, where the (M,N) output space is
divided equally across the thread blocks.

\subsubsection{Software pipelining and double buffering} \label{sec:software-pipeling}

Importantly, the Virgo-optimized GEMM kernel employs \emph{software pipelining}
which is enabled by the asynchronous programming interface of the matrix unit.
As shown in Figure~\ref{fig:pipeline}, while the matrix unit is computing a tile
along the K dimension consisting a \emph{consumer} pipeline, either the DMA unit
or a set of SIMT core warps collaboratively fetch the next input tile along
the K dimension from the global memory to the shared memory, consisting the
\emph{producer} pipeline. Another set of SIMT core warps can collaborate to form
an additional consumer pipe that does post-processing activation compute on a
previous result tile.  Because both the producer and consumer pipes run in
parallel, the tile data are double-buffered in the shared memory.  This
mechanism allows both the SIMT core warps and the matrix unit to participate in
useful work at all times, maximizing utilization of all hardware components in
the cluster.

\begin{figure}[t]
  \centering
  \includegraphics[width=\linewidth]{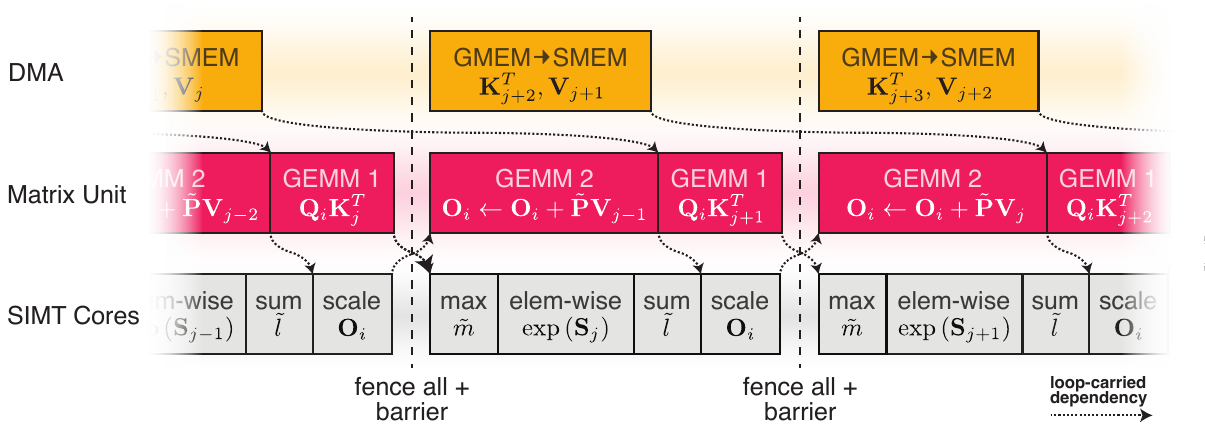}
  \caption{FlashAttention-3 mapping to Virgo.}
  \label{fig:flash-pipeline}
\end{figure}

\subsection{Mapping to FlashAttention-3} \label{sec:flash-mapping}

\hll{
To better illustrate the flexibility of Virgo's programming model, we
demonstrate how to map FlashAttention-3~\cite{shah2024flashattention3},
a GPU kernel that fuses the self-attention layer in the Transformer
architecture, to Virgo.  Detailed utilization and energy measurements for this
kernel are provided in Section~{\ref{sec:eval-flash}}.
}

\hll{
FlashAttention-3 defines three compute pipelines that can be overlapped
to improve hardware utilization: (1) \emph{GEMM-1} of $\mathrm{Q}$ and
$\mathrm{K^T}$ tiles, (2) \emph{GEMM-2} of $\mathrm{P}$ and $\mathrm{V}$ tiles,
and (3) softmax.
In the Virgo microarchitecture, we can map the two GEMMs sequentially to the matrix
unit, and the softmax to the SIMT core, allowing both to execute concurrently.
Given the producer-consumer relationship between the GEMM and softmax pipelines,
we use fence calls and cluster-wide barriers to synchronize and enforce ordering
across compute units.  This enables a software-pipelined
mapping onto the hardware, along with double-buffering of
intermediate tiles in shared memory. Figure~{\ref{fig:flash-pipeline}}
illustrates this mapping scheme.
}
\hll{
Listing~{\ref{lst:flash}} provides simplified pseudocode of a Virgo kernel implementing
this scheme.  In the loop iterating over tiles, the SIMT threads
initiate matrix unit operations for the two GEMMs, and
DMA operations to preload tile data into shared memory for the next iteration.
Since all operations are asynchronous, threads can immediately proceed with the
softmax computation within the SIMT core.  Notably, because both the GEMM and softmax
pipelines access the same activation tile (\texttt{smem\_O}), an additional
fence is required.  This fence only blocks until the GEMM-2 operation
completes, without affecting subsequent operations.
}

\begin{lstlisting}[language=C++, columns=fullflexible,
label={lst:flash},
caption=Pseudocode of FlashAttention-3 on Virgo.]

for (int iter = 0; iter < tiles; iter++) {
  // block until all previous-iter async ops complete
  virgo_fence(0);
  // synchronize cores cluster-wide at every loop iteration
  threadblock_barrier();
  // double-buffer SMEM tile addresses
  smem_Pp = (iter%2) ? smem_P0 : smem_P1; // produce
  smem_Pc = (iter%2) ? smem_P1 : smem_P0; // consume
  ...
  // initiate asynchronous matrix ops
  // GEMM-2: O = O + P*V 
  virgo_compute(smem_Pc, smem_Vc, smem_O, /*accum=*/1);
  // GEMM-1: S = Q*K
  virgo_compute(smem_Qc, smem_Kc, smem_Sp, /*accum=*/0);
  // asynchronously copy K and V tile for the next iter
  virgo_dma_load(gmem_K, gmem_V, smem_Kp, smem_Vp, ...);
  ...
  // overlap with SIMT softmax compute
  online_softmax(smem_Sc, smem_Pp, per_row_scale, ...);
  // block until the 2 most recent operations (GEMM-2)
  virgo_fence(2);
  row_rescale(smem_O, per_row_scale);
}
...
// copy final activation tile to global memory
virgo_fence(0);
virgo_dma_store(gmem_O, smem_O);

\end{lstlisting}

\subsubsection{Synchronization Overhead} \label{sec:programming-sync-overhead}

\hll{
To maximize overlap and achieve high utilization in the mapping scheme
described, the synchronization between the core and the matrix unit
must have minimal overhead.
To assess this, we measure the cycle times the core spends in the busy register
polling loop within the \texttt{virgo\_fence} operation.
On average, this interval is 260 cycles, which accounts for 2.4\% of the total
runtime. This indicates that the kernel exploits high instruction-level
parallelism, as all remaining cycles overlap matrix operations with SIMT
computations.
}

\hll{
The synchronization overhead also depends on the pipelining scheme used for the
workload and its throughput-matching behavior.  In our experiment,
the softmax execution completes slightly earlier than the GEMM. In
workloads where pipeline throughputs are ideally matched, or where the matrix
computation completes first, the synchronization overhead can be lower than what
we measure.
}

\section{Methodology} \label{sec:methodology}

 In this section, we outline how we implement both our proposed
 microarchitecture and the baseline GPU designs in RTL, and then set up
 experiments for evaluation.

\subsection{Baseline GPU Designs} \label{sec:method-baseline}

We first describe how we implement the baseline GPU designs that feature
core-coupled integration methods of matrix unit.
We continue to follow the classification we established in
Section~\ref{sec:terminology}, and focus on how to faithfully model each of the
design points.  This allows us to better evaluate our design in a way that
achieves separation of concerns without conflating multiple design
considerations at play.


\subsubsection{Volta-style Tightly-coupled Design}

The Volta-style design features matrix units tightly coupled to the core's
register file. For its implementation, we closely refer to the microarchitecture
proposed in~\cite{raihan2019modeling}, whose timing behavior is correlated with
the NVIDIA GPU. The microarchitecture features SIMD-parallel dot-product units,
composed of floating-point units in a tree-reduction configuration, that execute
multiply-and-accumulate (MAC) operations on the input tile fragments.  As a
result of tight coupling, the tile fragments are directly supplied from the
register file via core pipeline, and the result fragments are written back to
the register file as well.

We extend the Vortex RISC-V ISA to control the matrix unit, modeling Volta's
HMMA SASS instructions. Specifically, a single tile operation
is split into finer-grained \emph{set} and \emph{step} instructions, sequenced
to perform the inner and outer products of the input tile fragments.
\hll{It is worth noting that the Ampere and Hopper Tensor Cores manage the sequencing
of sets and steps at the microarchitecture level~\cite{raihan2019modeling, sun2022dissecting}, reducing
the overall instruction count.  We model this mechanism in the Hopper-style
baseline as well.}




Since our Vortex core configuration has a narrower SIMT width than Volta
(8 vs. 32), we correspondingly scale the design to achieve full
efficiency of the hardware. Specifically, we instantiate a single octet instance
instead of four in the original design~\cite{raihan2019modeling}.
The tile size of a single \verb|wmma| instruction is determined
by the register file space.  As discussed in Section~\ref{sec:accum-mem}, Tensor
Core operates at a per-warp granularity, and therefore can access the
register space SIMT-privatized to each warp, which is 1 KiB in our configuration.
This space allows storing two 8\texttimes16 FP16 operands and one 8\texttimes8
FP32 accumulator tile, resulting in the tile size of \verb|(m,n,k)|=\verb|(8,8,16)|.
The available read bandwidth from the register file limits the number of FP16 MACs per unit
to 32.
%
\hll{Our resulting Tensor Core implementation maintains the same timing
behavior of 2 cycles per \texttt{HMMA} \emph{step} instruction as in Volta, along with an
identical Tensor-to-SIMT FLOPS ratio of 4:1.
}

\subsubsection{Ampere-style Tightly-coupled Design with DMA}

\hll{The Ampere-style baseline features tightly-coupled matrix units and a dedicated
DMA incorporated into the GPU.  To model this baseline, we use an identical
matrix unit module as the Volta-style baseline, but additionally
instantiate a DMA unit described in Section}~\ref{sec:dma}.

Note that publicly available information does not explicitly confirm whether the
Ampere architecture includes a dedicated DMA engine similar to
Hopper~\cite{nvidia2024ampere}. However, to more clearly isolate the impact of
different design components in our evaluation, we introduce DMA in the
Ampere-style design and incrementally incorporate operand decoupling in the
Hopper-style design.

\begin{figure}[t]
  \centering
  \includegraphics[width=\linewidth]{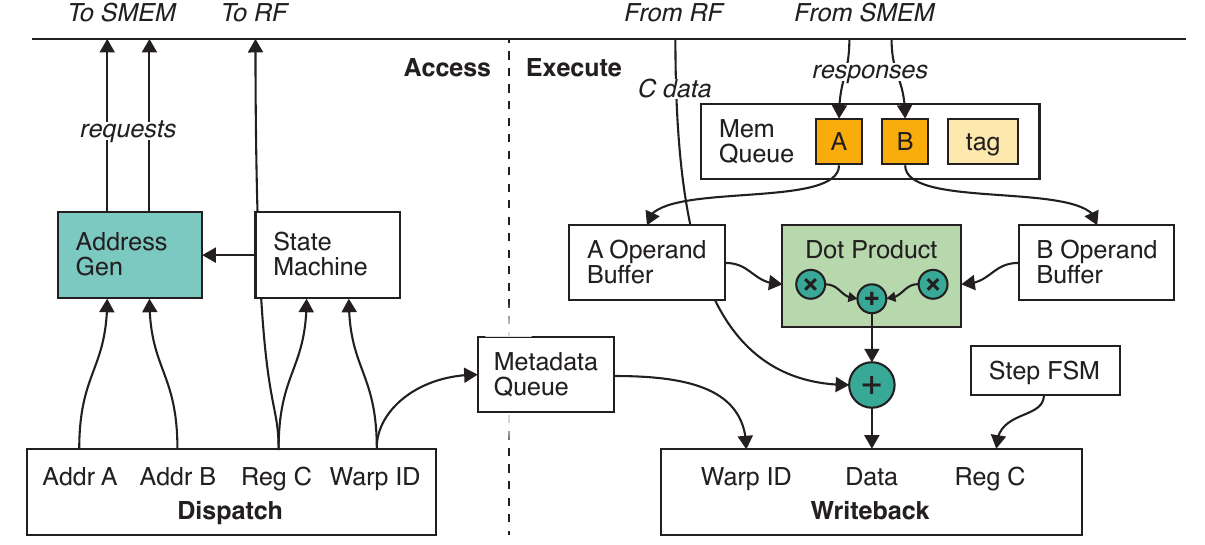}
  \caption{\hll{Microarchitecture of the decoupled Tensor Core, which models
  NVIDIA Hopper's integration scheme of decoupling operand storage to the shared
  memory~\cite{choquette2023hopper}.  The decoupled access/execute architecture
  effectively overlaps the access latency to the shared memory with compute.
  }}
  \label{fig:tensor-hopper-uarch}
\end{figure}

\subsubsection{Hopper-style Operand-decoupled Design} \label{sec:method-hopper}

\hll{
The Hopper-style baseline enhances the matrix unit design by enabling direct
access to operand matrix data from shared memory, while accumulator
data continues to be both read from and written to the
register file~\cite{nvidia2024ptx}. To implement this design, we extend the
Volta-style matrix unit implementation into a decoupled access-and-execute
architecture~\cite{smith1982decoupled}, illustrated in
Figure~{\ref{fig:tensor-hopper-uarch}}.  The \emph{access} frontend includes a
state machine and an address generation module that issues a sequence of read
requests to the matrix fragments stored in the shared memory. The \emph{execute}
backend comprises decoupling FIFO queues and operand buffers, which stage the
responses from shared memory and perform MAC operations when operand data
becomes available.
Since the addresses of the matrix fragments are statically
determined, the access frontend can run ahead of the execute backend,
effectively hiding shared memory access latency.

We expose an asynchronous instruction interface for the Hopper-style through
ISA extension that models the \texttt{wgmma} PTX
instruction~\cite{nvidia2024ptx}.  A warp kicks off the matrix unit via an
\emph{initiate} instruction without stalling, and later executes a \emph{wait}
instruction to synchronize with the unit and retrieve a valid result.  Similar to
Virgo, this asynchronous design enables instruction-level parallelism within a
single warp.

While offloading operand data alleviates capacity constraints, storing
accumulator data in the register file continues to limit tile size.
Similar to Volta, the Hopper unit has access to 1 KiB of register space
per warp, accommodating up to a single 16\texttimes16 FP32 accumulator tile.
There is no significant constraint to the shared-memory-stored operand tile
size; however, a large \texttt{k} tile dimension is not necessarily advantageous
as there is no data reuse over \texttt{k}. Consequently, the maximum
tile size we support in the Hopper unit is \texttt{(m,n,k)}=\texttt{(16,16,32)}.
The number of MACs in the unit is constrained by the shared memory
bandwidth, which is 64 FP16 operations per cycle in our configuration. This
achieves the Tensor-to-SIMT FLOPS ratio of 8:1.

Note that the Hopper-style design also incorporates a dedicated DMA, similar
to Ampere-style and Virgo.
%
}

\begin{table}[tb]
  \small
  \centering
  \begin{tabular}{c|c}
    \toprule
      \multicolumn{2}{c}{GPU SoC Configuration} \\
    \midrule
      Clusters  & 1 \\
      Cores per cluster  & 8 in Volta-style / 4 in Hopper-style \\
      SIMT Width& 8 warps/core, 8 lanes/warp \\
      SIMD Units & 2 ALUs, 1 FPU, 32 LSQ entries per lane \\
      Register File & 8KB INT, 8KB FP per core \\
      Shared Memory & 128 KB, 4 banks, 8-16 sub-banks \\
      Cache & 16KB L1I, 16KB L1D per core, 512KB L2 \\
    \midrule
      \multicolumn{2}{c}{Volta and Ampere-style Tightly-coupled Matrix Unit} \\
    \midrule
      \# of Units & 1 per core, 8 per cluster \\
      MACs (Tensor:SIMT) & \hll{32 FP16 (4:1)} / 16 FP32 (2:1) \\
    \midrule
      \multicolumn{2}{c}{\hll{Hopper-style Operand-decoupled Matrix Unit}} \\
    \midrule
      \# of Units & 1 per core, 4 per cluster \\
      MACs (Tensor:SIMT) & \hll{64 FP16 (8:1) / 32 FP32 (4:1)} \\
    \midrule
      \multicolumn{2}{c}{Virgo Disaggregated Matrix Unit} \\
    \midrule
      \# of Units & 1 per cluster \\
      Systolic Array & \hll{16 $\times$ 16 FP16} / 8 $\times$ 8 FP32 \\
      Accumulator Memory & 32KB \\
    \bottomrule
  \end{tabular}
  \caption{Hardware configuration of the GPU designs used for evaluation.}
  \label{tab:hw-configuration}
\end{table}

\subsection{System Implementation}

We leverage and extensively modify several key pieces of open-source hardware
infrastructure to enable our full system at the RTL level. At the top, we use
the Chipyard SoC generator framework~\cite{amid2020chipyard} to integrate all
hardware components into a single SoC design: Vortex SIMT cores, matrix units,
shared memory, cache, interconnects, etc. Parameters such as the number of cores
and clusters, the size and data type of the matrix unit, as well as memory bus
widths can all be adjusted, allowing Virgo to be a flexible and comprehensive
generator covering a large design space.

\textbf{Vortex SIMT Cores.} We take the SIMT core module from Vortex~\cite{tine2021vortex}
(\verb|vx_core|) to construct our own cluster hierarchy and synchronization
mechanisms, rather than incorporating the full top-level design. We also adapt
Vortex's cache bank for our L1 cache, positioned after the coalescer.


We made changes to the core pipeline to facilitate higher instruction and memory throughput. These include store fencing, optimized warp scheduling, synchronization, among others. Finally, we added Tensor Cores to the pipeline.


\textbf{Gemmini-based Matrix Unit.} We make extensive modifications to
Gemmini~\cite{gemmini-dac} in order to make the systolic array directly
interface with the cluster-level shared memory. Gemmini coordinates the movement
of the source matrices from its private wide scratchpad banks through its
systolic array, accumulating at either the individual processing elements or at
its private accumulator memory, depending on the selected dataflow. For Virgo,
we retained the accumulator memory to ensure full systolic array throughput,
which required single-cycle accumulation. We modified the scratchpad interface
to use the TileLink protocol and go through the SMEM interconnect, instead of
directly interfacing with SRAMs. The SIMT cores can therefore deposit data into
the SMEM directly for Gemmini to process.





\subsection{Experiments} \label{sec:experiments}
In this subsection, we will detail the environment in which the workloads are
run, the specifics of the kernels, as well as the tools we use to run them.

\textbf{Hardware Setup.} We summarize in Table~\ref{tab:hw-configuration} the
hardware configuration of the GPU designs we evaluate, with the GPU SoC
configuration being shared by all. Both the Tensor Core and Virgo matrix unit
uses fully-pipelined 1 op/cycle \emph{hardfloat}~\cite{hauser2019hardfloat}
units to implement floating point MACs. All configurations possess the same
number of MACs per cluster for a fair comparison.

\textbf{Workloads.}  We implement GEMM kernels independently optimized for Virgo
and baseline designs, with different sizes: \hll{
256\texttimes 256\texttimes 256, 512\texttimes 512\texttimes 512, and
1024\texttimes 1024\texttimes 1024, all of which are stored in FP16.  We write
the kernels in C++ and use Vortex's LLVM compiler toolchain to generate
binaries.
}

\hll{To demonstrate how the SIMT cores can effectively collaborate with the
cluster-level matrix unit in a realistic workload, we implement the
aforementioned FlashAttention-3~\cite{shah2024flashattention3} onto FP32
configurations of Virgo and Ampere-style baseline designs. We
evaluate the forward pass with a sequence length of 1024, head dimension of 64
with a single head, and a batch size of 1. As the Vortex core lacks a
multi-function unit that accelerates exponential operations, we use a 2nd-order
Taylor approximation for exponentials, an optimization found in other literature
as well~\cite{zhang2024hedgehog, arora2024simple}. }



\begin{figure}[t]
  \centering
  \includegraphics[width=\linewidth]{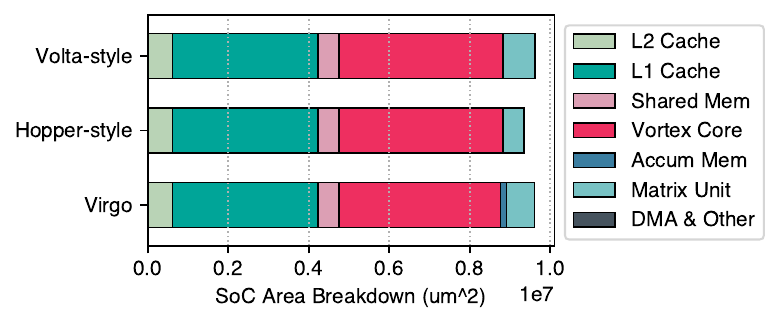}
  \caption{SoC area breakdown of the evaluated GPU designs.}
  \label{fig:soc-area-breakdown}
\end{figure}

\textbf{Tools \& Measurement.} We synthesize the design at 400 MHz using a
commercial 16 nanometer process. We use Cadence Joules to estimate power, and
use Cadence Genus to estimate area, as well as to verify the operating frequency.
An area breakdown of the synthesized SoC design configured to
Table~\ref{tab:hw-configuration} is shown in
Figure~\ref{fig:soc-area-breakdown}.  The Virgo design achieves 0.1\% smaller SoC area
than the Volta-style design, and 3.0\% larger area than the Hopper-style design.
The L1 cache consumes a large area as it is synthesized as flop arrays rather
than SRAMs; however, this did not affect the key findings in our evaluation.
Additionally, we used FireSim~\cite{karandikar-firesim-isca18} to verify the
design to be functional on a Xilinx Alveo U250 FPGA.

For all power and energy evaluation in the subsequent sections, we plot and
discuss \emph{active power}. This metric is obtained by taking the nominal SoC
package-wide power, and subtracting the power usage when it's fully idle. This
is necessary to provide a fair comparison, as the baseline idle power is highly
implementation-dependent.
For instance, we find the power consumption of the Vortex core to be very high
at idle when synthesized as ASIC; this is potentially because the core design
targets the FPGA environment, with common use of flop arrays and lack of clock
gating. By measuring the active power over idle, we can accurately characterize
the power implications of our design distinctly from the incidental
inefficiencies in the core implementation.


\section{Evaluation} \label{sec:eval}

We now discuss how Virgo's disaggregated integration leads to power and energy
reduction and utilization improvement using quantitative measurements from RTL
simulation and synthesis.  We focus on two metrics: power/energy
consumption and MAC hardware utilization, across two workloads: GEMM and FlashAttention-3~\cite{shah2024flashattention3}.

\subsection{GEMM Kernels} \label{sec:eval-gemm}

We first evaluate the GEMM kernels on both the cluster-integrated Virgo
design and the three baseline core-coupled designs, as described in
Section~\ref{sec:method-baseline}. Virgo achieves higher MAC
utilization across all GEMM configurations. Virgo is also
up to 67.3\% more power efficient and
80.3\%
energy efficient than the
Ampere-style core-coupled design, as well as 24.2\% more power efficient and
32.5\% more energy efficient compared to the Hopper-style operand-decoupled
design. We elaborate on the performance, power and energy advantage below.

\begin{table}[t]
    \centering
    \small
    \begin{tabular}{c|c|c|c}
    \toprule
      & \multicolumn{3}{c}{MAC \% Utilization} \\
     & 256$\times$256$\times$256 & 512$\times$512$\times$512 & \hll{1K$\times$1K$\times$1K} \\
     \midrule
         Volta-style        & 25.6          & 30.3          & \hll{30.3} \\
         Ampere-style       & 37.5          & 45.6          & \hll{52.3} \\
         \hll{Hopper-style} & \hll{60.5}    & \hll{72.8}    & \hll{77.0} \\
         \textbf{Virgo}     & \textbf{66.1} & \textbf{77.9} & \hll{\textbf{86.5}} \\
     \bottomrule
    \end{tabular}

    \caption{MAC unit \% utilization of the GEMM kernel, compared across the
      GPU designs with tightly-coupled matrix units (\emph{Volta-style}),
      tightly-coupled matrix units with DMA (\emph{Ampere-style}),
      \hll{operand-decoupled matrix units (\emph{Hopper-style})}, and Virgo.}
    \label{tab:gemm-perf}
\end{table}

\subsubsection{GEMM Performance} \label{sec:gemm-performance}

\hll{ Table~{\ref{tab:gemm-perf}} lists the cycle count and utilization figures
for the three GEMM sizes.
As shown in the table, the cluster-level matrix unit in Virgo achieves improved
MAC utilization compared against all three baselines described earlier.
%
%

Virgo achieves higher utilization mainly by enlarging the granularity of
operation, and thereby reducing the number of additional instructions that need
to be executed in the core pipeline. The Volta-style tightly-coupled Tensor
Core is driven by fine-grained instructions with a smaller tile size of
8\texttimes8, which means the SIMT cores need to execute more Tensor Core
instructions to complete the full GEMM.  Moreover, the input matrix elements
need to be loaded into the register file via load instructions from the shared
memory, each of which is also preceded by additional address generation
instructions.  This overall increased number of instructions leads the matrix
unit utilization to become constrained by the instruction throughput available
from the SIMT core. We note that such constraint can be partially alleviated by
microarchitecture improvements in the core, such as dual-issue of instructions.

In contrast to Volta and Ampere, the Hopper-style operand-decoupled Tensor Core
significantly reduces instruction overhead within the core through three key
factors.
First, by increasing the tile size to 16\texttimes16\texttimes32 (Section~\ref{sec:method-hopper}),
it lowers the number of matrix operations issued for a given GEMM dimension.
Second, unlike the fine-grained, synchronous Volta-style instructions, the unit
is driven by a microarchitectural state machine~\cite{raihan2019modeling,
sun2022dissecting}, further reducing kernel instructions.
Third, similar to Virgo, it autonomously retrieves operand data from shared
memory, eliminating load instructions in the kernel. Consequently, the
Hopper unit achieves significantly higher utilization than the tightly-coupled
baselines.

In a similar manner, Virgo further increases the tile size to
128\texttimes64\texttimes128, minimizing MAC utilization constraints imposed by the
core's available instruction throughput.

Instruction statistics support this
observation: the number of retired instructions in Virgo is only 0.5\% of those in
the Volta-style Tensor Core design and 8.0\% of those in the Hopper-style
operand-decoupled Tensor Core design.


}



\begin{figure}[t]
  \centering
  \includegraphics[width=\linewidth]{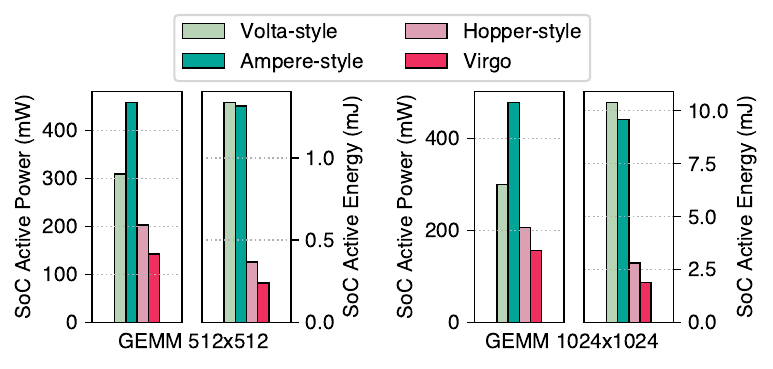}
  \caption{Active power and energy comparison between Virgo and the three baseline
  core-coupled designs.}
  \label{fig:soc-power-energy-512}
\end{figure}




\subsubsection{GEMM Power \& Energy} \label{sec:gemm-power-energy}

In Figure~\ref{fig:soc-power-energy-512} we show a side-by-side comparison
between the power usage of Virgo and the three Tensor Core baselines
running two GEMM sizes. Virgo enables a \textbf{67.3\%} and \textbf{24.2\%}
reduction in active power compared to the Ampere and Hopper-style baselines,
respectively. The overall higher power consumption of the Ampere-style design
over Volta-style may initially seem unintuitive, but is explained by the higher
hardware utilization, enabled by DMA via better operand delivery.



\begin{figure}[t]
  \centering
  \includegraphics[width=\linewidth]{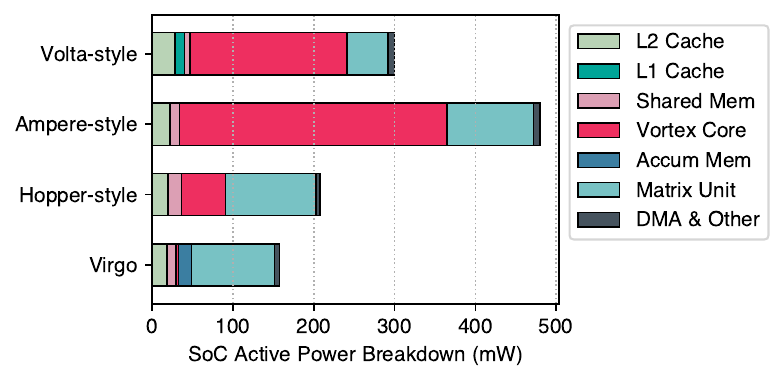}
  \caption{Active power breakdown by SoC components running 1024\texttimes1024
  GEMM.  The core active power is reduced significantly as a result of reduced
  instruction processing and register file accesses in Virgo.}
  \label{fig:soc-power-breakdown}
\end{figure}

\begin{figure}[t]
  \centering
  \includegraphics[width=\linewidth]{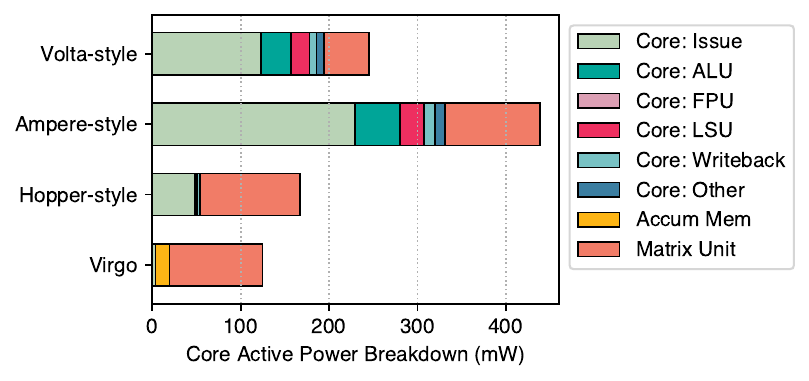}
  \caption{Active power breakdown within the Vortex SIMT core running
  1024\texttimes1024 GEMM.  The Virgo matrix unit component is included for
  comparison, although it resides outside of the SIMT core.}
  \label{fig:core-power-breakdown}
\end{figure}

\begin{figure}[t]
  \centering
  \includegraphics[width=\linewidth]{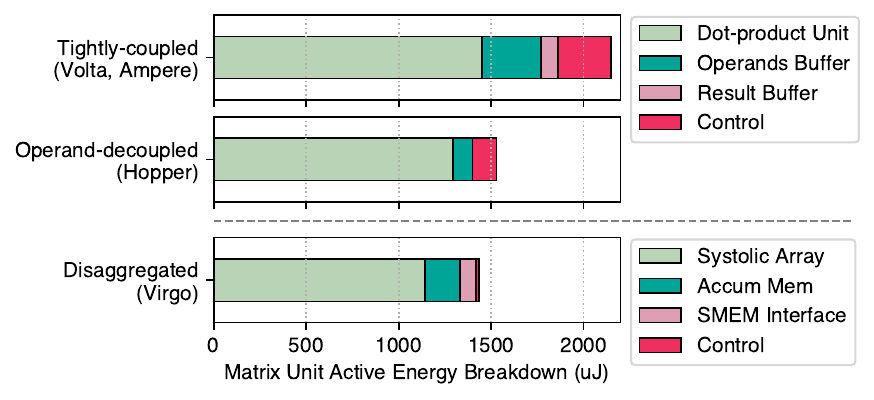}
  \caption{Active energy breakdown of the matrix unit for the 1024\texttimes1024
   GEMM kernel.}
  \label{fig:matrix-unit-energy-breakdown}
\end{figure}

To better reason about the source of power consumption in the hardware, we give
a detailed power breakdown at the SoC (Figure~\ref{fig:soc-power-breakdown}),
the SIMT core (Figure~\ref{fig:core-power-breakdown}), and energy breakdown at
the matrix units (Figure~\ref{fig:matrix-unit-energy-breakdown}).  At the SoC
level, the Vortex SIMT core is a major component in the overall power
consumption. Importantly, comparing Virgo to the Tensor Core designs reveals
that most power reduction occurs within the core rather than in the matrix units
themselves. We now outline the key factors contributing to such power savings.

\textbf{Larger operation granularity and reduced instructions.}
%
Figure~\ref{fig:core-power-breakdown} highlights a major difference in core
power consumption, particularly in the issue and ALU stages, when comparing
Virgo and Hopper-style designs to Volta/Ampere-style designs.
This significant power reduction is primarily attributed to the substantial
decrease in the number of kernel instructions, as explained in
Section~\ref{sec:gemm-performance}.
Executing fewer instructions in the core reduces energy consumption associated
with dynamically dispatching the instructions via the scoreboard and the warp
scheduler, which is reflected in the issue stage power.
Additionally, the larger tile sizes in Virgo and Hopper decrease the number of
address generation instructions required for the base memory address of each
tile, contributing to lower ALU energy consumption.  Collectively, Virgo's larger
operation granularity minimizes incidental power consumption in the hardware
logic outside the matrix unit.

\hll{
\textbf{Offloading operand access from the register file.} Beyond instruction
processing, accessing the register file for operand delivery also significantly
impacts core power.  In a Volta-style tightly-coupled Tensor Core, operand
matrices are accessed by fetching register operands from the register file as
encoded in the instruction.
Staging the operands through the register file incurs additional energy compared
to accessing them directly from the shared memory, as both the Hopper-style and
Virgo matrix units do.
This design choice explains the significant reduction in the issue stage power,
which includes register file access power, from the Ampere-style to Hopper and Virgo
figures.
}

\hll{\textbf{Decoupled accumulation.}
Although the operands are offloaded, the Hopper-style operand-decoupled Tensor
Core continues to access accumulator data from the core register file.   This
explains the nontrivial core issue stage power consumption in the Hopper-style
baseline (Figure~{\ref{fig:core-power-breakdown}}). In contrast, Virgo takes an
additional step by offloading accumulator data from the register file to a
dedicated SRAM-based accumulator memory.
The accumulator memory is single-banked, unlike the multi-banked register file
that needs to support divergent SIMT accesses, reducing energy consumption per
access. In addition, the Virgo matrix unit's systolic array architecture allows
accumulation to partially occur within the mesh, reducing the number of accesses
to the memory and further saving power.
}






\hll{
\textbf{Energy consumption in matrix units.}
Figure~{\ref{fig:matrix-unit-energy-breakdown}} shows the active energy
consumption of the matrix unit in isolation. We focus on energy rather than
power for the matrix unit, because energy directly correlates with the total
FLOPs which is kept the same across the designs. The figure does indicate that
energy consumption in the matrix unit, especially within the processing elements
(``Dot-product Unit'' and ``Systolic Array''), is similar across the designs.
The PE energy is slightly lower for Virgo, due to its use of fused multiply-add
units in the systolic array, which are more energy-efficient than the separate
multipliers and adders used in Tensor Core’s tree-reduction
PEs~\cite{raihan2019modeling}.
However, this difference in the matrix unit energy is minuscule compared to the
overall system-level energy savings at the SoC
(Figure~\ref{fig:soc-power-energy-512}).
This suggests that Virgo's advantages in power and energy efficiency primarily
come from the aspects of operation granularity and operand/accumulator
offloading, rather than from choosing a specific design for the matrix unit.
}


\begin{table}[t]
    \centering
    \small
    \resizebox{0.485\textwidth}{!}{
    \begin{tabular}{c|c|c|c|c}
    \toprule
       & Tile   & Matrix Unit & \multicolumn{2}{c}{SMEM Footprint} \\
       \cline{4-5}
       & Fragment      & Design      & \makecell{MiB} & \makecell{Norm.} \\
    \midrule
        Tightly-coupled   & 8\texttimes8           & Per-core    & 6 & 2.67 \\
        Operand-decoupled & 16\texttimes16          & Per-core & 4 & 1.78 \\
        \textbf{Disaggregated} & \textbf{16\texttimes16} & \textbf{Per-cluster} & \textbf{2.25} & \textbf{1.00} \\
    \bottomrule
    \end{tabular}
    }

    \caption{\hll{On-chip shared memory read footprint compared across the GPU
    designs.  Virgo's disaggregated matrix unit improves data reuse by (1)
    enlarging tile size over the tightly-coupled design, and (2) unifying into
    a single cluster-level unit over the operand-decoupled design.}}
    \label{tab:smem-footprint}
\end{table}

\subsubsection{Data Reuse in the Shared Memory} \label{sec:eval-smem-reuse}

\hll{
Table~{\ref{tab:smem-footprint}} lists the shared memory read footprint for the
256{\texttimes}256\texttimes 256 GEMM kernel.
%
``Tile fragment'' refers to the amount of matrix data transferred from shared
memory to compute units in a single access.
Its size depends on the backing memory internal to the matrix unit that stages
data between shared memory and PEs: the operand buffer for both tightly-coupled
and operand-decoupled Tensor Cores, and the systolic array registers for
Virgo. Unlike the tightly-coupled design, where tile size is
severely constrained by the register file space, operand-decoupled and
disaggregated designs increase tile fragment size by offloading operand storage,
improving data reuse as seen in the lower footprint.

More importantly, Virgo further improves data reuse beyond the operand-decoupled
design by unifying the matrix units into a single instance at the cluster level.
This unified design enables data sharing in scenarios where non-unified Tensor
Cores would otherwise duplicate shared memory accesses.
Specifically, in GEMM, different Tensor Cores may compute output tiles along the
same row or column. However, due to their physical separation across cores, they
must independently access the input row or column data multiple times.
In contrast, a unified design allows complete reuse of input matrix data,
which explains the further memory footprint reduction for Virgo compared to the
Hopper-style operand-decoupled design.



This reduction in memory footprint has both performance and energy implications.
We had to scale up the shared memory bandwidth by 2x for Volta and Ampere-style
designs with a more aggressive banking strategy, because otherwise
the matrix unit utilization would be bottlenecked by memory bandwidth.
In one configuration, this change increased utilization from 46.9\% to 55.0\%.
With better data reuse, Virgo achieved optimal utilization without demanding
higher memory bandwidth. Energy-wise, the shared memory in the Virgo design
consumed 41.0\% less active energy compared to the operand-decoupled design. }


\begin{figure}[t]
  \centering
  \includegraphics[width=\linewidth]{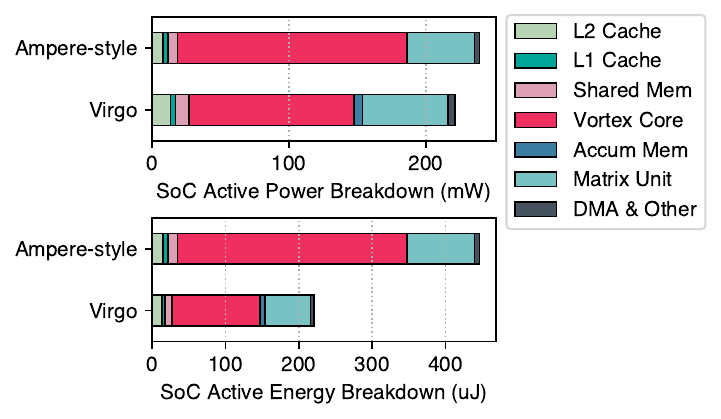}
  \caption{FlashAttention-3 active power and energy breakdown.}
  \label{fig:flash-soc-power-breakdown}
\end{figure}

\subsection{FlashAttention-3 Kernel} \label{sec:eval-flash}

\hll{We perform a similar analysis of performance, power and energy for
the FlashAttention-3 kernel~\cite{shah2024flashattention3}, whose mapping to Virgo
is described in Sections~{\ref{sec:flash-mapping}} and
{\ref{sec:experiments}}. We compare Virgo against the Ampere-style design with
tightly-coupled matrix units and an integrated DMA.
For the Ampere-style design, the kernel mapping adopts the warp specialization
and ping-pong scheduling approach detailed in~\cite{shah2024flashattention3},
where the 8 warps within a core are divided into two groups of 4, with GEMM and
softmax computations scheduled alternately across the two warp groups.
Although the Ampere-style matrix unit lacks single-warp asynchronous execution,
it can still co-execute alongside SIMT vector units across different warps,
provided the warp scheduler can find eligible warps to concurrently
multi-thread.
%
}

\hll{\textbf{Performance \& Utilization.} We observe a total MAC utilization of
65.7\% for Virgo, a significant increase over 35.1\% for the Ampere-style
baseline.}
\hll{
The higher utilization is explained by the disaggregated design enabling more
efficient allocation of core pipeline resources to non-matrix computations.
Since matrix computations are offloaded to a disaggregated hardware, a single
warp, after asynchronously initiating the GEMM operation, can reallocate all of
its pipeline resources such as instruction slots and register file space to
softmax computation.
In contrast, the tightly-coupled matrix unit operates through synchronous,
fine-grained instructions, meaning that both matrix and softmax instructions
need to compete for instruction issue slots and limited register file ports at
every cycle.  This leads to suboptimal parallelism compared to the disaggregated
design.
}

\hll{\textbf{Power \& Energy.} We plot the active SoC-level power and energy
consumptions of the two designs in Figure~{\ref{fig:flash-soc-power-breakdown}}.
As a result of fusing matrix operations with SIMT operations, the core power
constitutes more than 50\% of the total power for both designs. However, due to
the larger operation granularity and offloading of data delivery from the
register file as discussed in Section~{\ref{sec:gemm-power-energy}}, the core
power consumption is reduced for Virgo, substituted with the matrix unit power
operating at a higher utilization. Compounded with the shorter overall runtime
of the kernel, the overall energy consumption of the kernel is reduced by 50.6\%
compared to the Ampere-style baseline.}

\subsection{Multiple Heterogeneous Matrix Units} \label{sec:multi-gemmini}

The disaggregation of matrix units and parameterized memory system of Virgo
allows scaling to multiple matrix units within a cluster. We showcase a
configuration where we integrate two differently sized matrix units in one cluster,
and map two differently sized GEMMs on them: a 256\texttimes 256\texttimes 256
GEMM onto the large unit, and a 128\texttimes 128\texttimes 128 GEMM onto the
small unit. When the two GEMMs run in parallel, we achieve a utilization of
59.5\%, compared to 59.7\% when they are run separately in serial. The active
power when normalized by FLOP increased by only 4.3\% for the parallel case. The
minimal performance degradation and power overhead demonstrates the scalability
of Virgo's design.

\section{Conclusion}
We introduce Virgo, a novel GPU architecture that integrates matrix units
at the cluster level. By physically disaggregating the matrix unit from the core
pipeline, Virgo overcomes scalability and energy efficiency
limitations in the current core-coupled GPU designs.
This approach enables a heterogeneous GPU architecture optimized for DNN
workloads, where SIMT cores and matrix hardware are decoupled for improved
scalability while also supporting concurrent execution through efficient command and
memory coordination.



We implement and evaluate our design using synthesizable RTL, which shows
significant power and energy efficiency improvement for GEMM operations.
Furthermore, the decoupled architecture in conjunction with our programming
model enables efficient mapping to fused DNN kernels, enabling new strategies for
applications mapping.

\begin{acks}
We would like to thank our shepherd Joseph L. Greathouse, the anonymous
reviewers and artifact evaluators for their valuable feedback.  We would also
like to thank Amir Gholami for providing access to a server with NVIDIA GPUs
that we used to characterize GEMM kernels.  Lastly, we extend our sincere
thanks to Kostadin Ilov for his management of the compute servers we used for
development.

This research was partially funded by NSF Awards 2303735 and 2238346, as well as
by SLICE Lab industrial sponsors and affiliates.  The views and conclusions
contained in this document are those of the authors and should not be
interpreted as representing the official policies, either expressed or implied,
of the U.S. Government.
\end{acks}

\clearpage

\appendix
\section{Artifact Appendix}

\subsection{Abstract}

We evaluate the artifacts of this paper via RTL simulation of the
implementation of our proposed GPU microarchitecture, executing
two main workloads: GEMM and FlashAttention-3 kernels.
We use cycle numbers from the RTL simulation to evaluate hardware
utilization, energy measurements, and memory footprint metrics discussed
in the paper.

\subsection{Artifact check-list (meta-information)}


{\small
\begin{itemize}
  \item {\textbf{Compilation:} GPU kernel compiled with Vortex LLVM compiler, RTL elaborated using Chisel and Chipyard. }
  \item {\textbf{Hardware:} Ubuntu 20.04 server with Synopsys VCS installed. }
  \item {\textbf{Metrics:} Cycle time from the RTL simulation, power and energy consumption, MAC hardware utilization, memory footprint. }
  \item {\textbf{Output:} Microarchitectural trace log, cycle reports in CSV, utilization numbers, plots in PDF.}
  \item {\textbf{Experiments:} RTL simulation of GPU hardware executing GEMM and FlashAttention-3 kernels.}
  \item {\textbf{How much disk space required (approximately)?:} 150GB}
  \item {\textbf{How much time is needed to prepare workflow (approximately)?:} 30 mins }
  \item {\textbf{How much time is needed to complete experiments (approximately)?:} 30 hours }
  \item {\textbf{Publicly available?:} Yes}
  \item {\textbf{Code licenses (if publicly available)?:} BSD-3, Apache}
  \item {\textbf{Archived (provide DOI)?:} Planned }
\end{itemize}
}

\subsection{Description}

As discussed in Section~\ref{sec:ae-dependencies}, our RTL
implementation requires the Synopsys VCS compiler for correct
and reproducible simulation.  The following setup instructions
need to be carried out on a Linux compute server that has
access to a working VCS installation.  Aside from VCS,
all code and peripheral data will be set up from scratch.

\subsection{Installation}

\subsubsection{How to access}

Obtain the artifact tarball from \url{https://doi.org/10.5281/zenodo.14835068}.
The artifact consists of:

\begin{itemize}
\item{Chisel RTL implementation of the Virgo design,}
\item{GPU kernels evaluated on the Virgo implementation,}
\item{Chipyard framework ~\cite{amid2020chipyard} that integrates IPs into an SoC,}
\item{Post-processing and plot generation scripts, and}
\item{Prebuilt Vortex~\cite{tine2021vortex} toolchain binaries.}
\end{itemize}

Please note that we provide up-to-date source code and instructions at the
public Github repositories \url{https://github.com/ucb-bar/virgo} and
\url{https://github.com/ucb-bar/virgo-kernels}.

%


\subsubsection{Extracting tarball}


In a Linux shell, run the following commands:

\begin{minted}[breaklines]{bash}
 curl -O "https://zenodo.org/records/14835069/\
   files/virgo-artifact-full.tar.gz?download=1"
 tar xzvf virgo-artifact-full.tar.gz
 cd virgo-artifact-full/
\end{minted}

We will refer to \texttt{virgo-artifact-full/} as the \emph{root directory} going
forward.



\subsubsection{Setting up Chipyard}
At the root directory, run the following:

\begin{minted}[breaklines]{bash}
 cd chipyard
 # this might take ~10 minutes
 ./build-setup.sh riscv-tools --skip-submodules \
   --skip-firesim --skip-marshal
\end{minted}


\subsubsection{Compiling Virgo kernels}
At the root directory:
\begin{minted}[breaklines]{bash}
 cd virgo-kernels
\end{minted}
Inside \verb|virgo-kernels|, \texttt{lib} contains the Vortex SIMT core runtime
and its header files. \texttt{kernels} contains the source code for the kernels.
Run the following to compile all kernels:
\begin{minted}[breaklines]{bash}
 source ./scripts/toolchain_env.sh
 # compile the Vortex runtime static library
 make -C lib
 # compile GEMM kernels for Virgo
 cd kernels/sgemm_gemmini_dma
 ./compile_virgo.sh
 # compile GEMM kernels for tensor cores
 cd ../sgemm_tcore
 ./compile_tcore.sh
 # compile FlashAttention kernels
 cd ../flash_attention
 ./compile_flash.sh
\end{minted}

\subsubsection{Compile simulation binaries}
Exit to the root directory, then:
\begin{minted}{bash}
 cd chipyard/sims/vcs
 source ../../env.sh
 # compile simulator binaries (~10 minutes)
 ./scripts/compile_designs.sh
\end{minted}

\subsection{Basic test}
To check if everything worked up till this point, under \\ 
\texttt{chipyard/sims/vcs}, run
\begin{minted}{bash}
 ./scripts/sanity.sh
\end{minted}

You should see this output:
\begin{minted}{text}
 Sanity check passed!
\end{minted}

\subsection{Experiment workflow} \label{sec:ae-workflow}

\subsubsection{Running RTL simulations}
Ensure you are in a \texttt{tmux} or a \texttt{screen} session, as this step
runs a long process. Under \texttt{chipyard/sims/vcs}:
\begin{minted}[breaklines]{bash}
 source ../../env.sh
 # start kernel runs
 ./scripts/run_sims.sh
\end{minted}
We estimate that the longest simulation, i.e. 1024\texttimes1024 GEMM on the
Volta configuration, will take around 24 to 30 hours. The script shows the
current simulation timestamp in nanoseconds and clock rate in Hz; you should see
a total of 14 concurrent simulations.

The subsequent steps require RTL simulation results. If you continue
before all simulations are finished, their outputs may be partial.

\subsubsection{Verifying hardware utilization} From
\texttt{chipyard/} \texttt{sims/vcs}, run:
\begin{minted}[breaklines]{bash}
 ./scripts/utilization.sh > \
   ../../../scripts/cycles.csv
 ./scripts/utilization.flash.sh > \
   ../../../scripts/cycles.flash.csv
\end{minted}

The script will process the microarchitectural trace logs generated from the RTL
simulations, and print to the terminal the MAC utilization numbers that should
match Table~\ref{tab:gemm-perf} for GEMM, and the ``Performance \& Utilization''
paragraph in Section~\ref{sec:eval-flash} for FlashAttention.

\subsubsection{Reproducing power/energy measurements and plots} \label{sec:ae-plots}
Back at the root directory outside \texttt{chipyard}, run:

\begin{minted}[breaklines]{bash}
 cd scripts/
 python run_plots.py
\end{minted}

The script will output reproduced metrics along with the directions of how to
link the metrics to figures and values in the paper.
Please note that if the previous step of running RTL simulations is not
finished, the generated plot might be missing some bars.  Specifically,
the 1024\texttimes1024 dimension GEMMs for Virgo and Ampere may take a long time (>30 hrs),
and running the script before their completion may result in empty energy plots.

Please also note that due to licensing restrictions with the commercial PDK
used for RTL synthesis, we do not provide methods for reproducing
power and area measurements in the paper.  Instead, we supply static CSV files,
\verb|power_*.csv| and \verb|area_summary.csv|, that contain the power
measurement data used as input for the post-processing scripts.
However, because the cycle latency from RTL simulation is dynamically evaluated,
the energy metrics will also be dynamically generated and evaluated by
multiplying the power measurement with the cycle latency.
\newline
\subsubsection{Verifying shared memory footprint metrics}~
\newline
From \texttt{chipyard/sims/vcs}, run:
\begin{minted}[breaklines]{bash}
 ./scripts/smem_util.sh
\end{minted}

The script will process the FSDB waveform files generated from the RTL simulation,
and report the access footprint to the shared memory that shall
match Table~\ref{tab:smem-footprint}.

\subsubsection{Software dependencies} \label{sec:ae-dependencies}

The Vortex toolchain binaries provided in the artifact are built for an Ubuntu
20.04 LTS system or newer. Additionally, we find that Verilator fails to
simulate our RTL due to crashes, and having access to the Synopsys VCS compiler
is necessary for correct simulation.  Our evaluation is tested
with VCS version V-2023.12-SP1.

The post-processing and plotting scripts require a working Python 3 environment
with following packages installed: \verb|ripgrep|, \verb|pandas|,
\verb|matplotlib|, \verb|numpy|.  These are used for generating input data for
the kernel workloads, parsing log outputs, and generating plots.  Setting up a
Conda environment with these packages installed may be helpful.

\subsection{Evaluation and expected results}

We explain how the reproduced results link to the contents of the paper
in the individual steps listed in Section~\ref{sec:ae-workflow}, and
also in the terminal outputs from the scripts.
With all of the above steps completed, we expect the following results
to have been evaluated:

\begin{itemize}
\item{Cycle execution latency and hardware utilization of the GPU
running GEMM and FlashAttention kernels (Table~\ref{tab:gemm-perf}
and Section~\ref{sec:eval-flash}),}
\item{Power and energy measurements and plots for GEMM and FlashAttention
(Section~\ref{sec:eval-gemm}, \ref{sec:eval-flash}, Figure~\ref{fig:soc-power-energy-512}-\ref{fig:matrix-unit-energy-breakdown})}
\item{Shared memory footprint measurements (Table~\ref{tab:smem-footprint}).}
\end{itemize}

Specifically, the plots generated in Section~\ref{sec:ae-plots} correspond
to the following figures in the paper:

\begin{itemize}
\item{Figure 8: \verb|fig-soc-power-energy-1x4.pdf|}
\item{Figure 9: \verb|fig-soc-power-breakdown.pdf|}
\item{Figure 10: \verb|fig-core-power-breakdown.pdf|}
\item{Figure 11: \verb|fig-matrix-unit-energy-breakdown.pdf|}
\item{Figure 12: \verb|fig-flash-power-energy-breakdown.pdf|}
\end{itemize}

%




\subsection{Methodology}

Submission, reviewing and badging methodology:

\begin{itemize}
  \item \url{https://www.acm.org/publications/policies/artifact-review-badging}
  \item \url{http://cTuning.org/ae/submission-20201122.html}
  \item \url{http://cTuning.org/ae/reviewing-20201122.html}
\end{itemize}

\clearpage


\bibliographystyle{ACM-Reference-Format}
\bibliography{references}

\AtEndDocument{\par\cleardoublepage}

\end{document}